\documentclass[twocolumn,twocolappendix]{aastex631}

\usepackage{threeparttable}%
\usepackage{mathtools}
\usepackage[T1]{fontenc}
\usepackage{tabularx}

\shorttitle{X-ray Observations of Binary SMBH Candidates}
\shortauthors{Breiding et al.}
\graphicspath{{./}{figures/}}

\begin{document}

\title{Chandra X-ray Observations of Quasars with Velocity-Offset Broad Lines: Assessing the Binary Supermassive Black Hole Hypothesis}

\correspondingauthor{Peter Breiding}
\email{pbreiding@gmail.com}

\author[0000-0003-1317-8847]{Peter Breiding}
\affiliation{Department of Physics, Applied Physics, and Astronomy, Binghamton University, Binghamton, NY 13902, USA}

\author[0000-0002-3719-940X]{Michael Eracleous}
\affiliation{Department of Astronomy \& Astrophysics and Institute for Gravitation and the Cosmos, The Pennsylvania State University, 525 Davey Lab, University Park, PA 16802, USA}

\author[0000-0002-7835-7814]{Tamara Bogdanovi\'c}
\affiliation{School of Physics and Center for Relativistic Astrophysics, 837 State St NW, Georgia Institute of Technology, Atlanta, GA 30332, USA}

\author[0000-0003-4052-7838]{Sarah Burke-Spolaor}
\affiliation{Department of Physics and Astronomy, West Virginia University, P.O. Box 6315, Morgantown, WV 26506, USA}
\affiliation{Center for Gravitational Waves and Cosmology, West Virginia University, Chestnut Ridge Research Building, Morgantown, WV 26505, USA}
\affiliation{Canadian Institute for Advanced Research, CIFAR Azrieli Global Scholar, MaRS Centre West Tower, 661 University Ave. Suite 505, Toronto ON M5G 1M1, Canada}

\author{T. Joseph W. Lazio}
\affiliation{Jet Propulsion Laboratory, California Institute of Technology, 4800 Oak Grove Dr, Pasadena, CA 91109, USA}



\begin{abstract}
	
During the final stages of a galaxy merger, dynamical friction acting on the supermassive black holes (SMBHs) in the post-merger remnant can lead to the formation of a gravitationally bound binary SMBH.  In the event that at least one of these SMBHs is actively accreting, the system can appear phenomenologically as an active galactic nucleus (AGN) with a broad line region (BLR) kinematically offset from the host galaxy rest frame. Such velocity offsets have been interpreted as signatures of binary SMBHs, recoiling SMBHs, or BLR gas dynamics within a single-SMBH system.  We present deep Chandra X-ray observations of five nearby ($0.1 < z < 0.2$) Sloan Digital Sky Survey quasars whose broad emission lines are Doppler-shifted relative to their host galaxies’ systemic velocities, along with archival Chandra observations of 11 additional sources from the same sample.  Using our Chandra data, we constrain SMBH masses with multiple independent techniques.  We find systematic, method-dependent differences among black hole mass estimates, with masses inferred from the fundamental plane of black hole activity generally lower and single-epoch virial masses typically higher than those obtained using other methods.  We also compare the X-ray photon indices and optical-to-X-ray spectral indices of our quasars to the broader quasar population.  While we find no strong differences in optical-to-X-ray spectral indices, we do find systematically harder X-ray photon indices than typically observed in comparable quasars.  These results constrain competing physical models but do not provide conclusive evidence for or against a binary SMBH origin of the velocity-offset BLRs.
	
\end{abstract}



\section{Introduction}\label{intro}

The hierarchical assembly of galactic stellar mass, galactic structure, and the mass of supermassive black holes (SMBHs) at the centers of galaxies is widely believed to be shaped by galaxy mergers \citep[e.g.,][]{kauffmann+00,hopkins+06}.  Gravitational torques generated during a galaxy merger can drive gas inflows toward the remnant’s center, potentially fueling star formation in the bulge, accretion onto the central SMBH, and triggering an active galactic nucleus (AGN) \citep[e.g.,][]{dimatteo+05,hopkins+06,ellison+19}.  Given that virtually all massive galaxies host central SMBHs \citep{kormendy_and_ho_13}, galaxy mergers inevitably bring two (or more) SMBHs into the gravitational potential of the merger remnant.  
In the scenario described originally by \cite{begelman80}, these SMBHs are progressively drawn toward the common center through dynamical friction and subsequent orbital evolution, eventually forming a bound binary whose inspiral is driven by gravitational-wave emission.  Such gravitational waves have recently been evidenced by pulsar timing arrays (PTAs) as a stochastic gravitational-wave (GW) background consistent with an ensemble of inspiraling binary SMBHs \citep[e.g.,][]{NANOGrav+23}.  The identification and characterization of binary SMBHs through traditional electromagnetic means will be essential for the multi-messenger science anticipated from future PTA observations \citep[][]{burke-spolaor+19,burke-spolaor+25}.  While gravitational-wave observations can identify candidate binary SMBHs, their sky localizations are often too coarse to securely associate the source with an individual galaxy.  Phenomenologically, the primary way to identify and study individual candidate binary SMBHs is through their manifestation as active galactic nuclei (AGN).

\subsection{X-ray Emission Expected in Binary SMBHs vs.  Single-SMBH AGN}

In this subsection we summarize X-ray signatures proposed primarily on the basis of theoretical and numerical studies; in \S\ref{intro:time_domain} we describe the observational time-domain and spectroscopic signatures most commonly used to identify candidate binary SMBHs.

X-ray emission provides one of the most direct probes of supermassive black hole (SMBH) accretion, originating from a hot, compact corona ($T \sim 10^{9}$~K) located above the accretion disk \citep[e.g.,][]{haardt+93,fabian+15}.  Although the coronal geometry, size, and energy-dissipation mechanism remain poorly constrained, a prevailing paradigm is that inverse-Compton scattering of optical/UV disk photons in this corona produces the observed power-law X-ray continuum.  A fraction of this continuum irradiates the inner accretion flow, which in turn can lead to fluorescent Fe~K$\alpha$ emission lines that encode the disk geometry and relativistic effects arising from gas orbiting close to the black hole event horizon \citep[see e.g.,  review by ][]{reynolds21}.

Numerical simulations suggest that binary SMBHs can sustain circumbinary disks plus persistent ``mini-disks'' around each black hole \citep[e.g.,][]{farris+14,duffell+20,munoz+20}.  In compact systems (with typical orbital separations a $\lesssim 100\,R_g$ in simulations, where $R_g \equiv GM/c^2$), the X-ray output may include (i) emission from the inner rim of the circumbinary disk, which may contribute thermal soft X-rays (Wien peak $\sim$1~keV) \citep[e.g.,][]{tang+18,dascoli+18}, (ii) shock-heated stream--disk impact regions (often described as transient ''hot spots'') that have been suggested in some simulations to produce hard X-rays up to tens--$\sim$100~keV \citep[e.g.,][]{roedig+14,tang+18,dascoli+18}, and (iii) emission from multiple mini-disks and their coronae, potentially adding spectral complexity and hardening relative to single-SMBH AGN \citep[e.g.,][]{tang+18,dascoli+18}.    At the smallest separations, the binary orbit can evolve rapidly due to gravitational-wave emission, so these features may be short-lived and not generically representative of wider-separation systems.

These considerations motivate X-ray diagnostics as a potential, though model-dependent and non-unique, probe of binary SMBHs. Numerical simulations predict soft X-ray emission below $\sim$1~keV from circumbinary material and a Compton reflection component peaking near $\sim$20~keV from the hottest regions of the mini-disks \citep[][]{ryan+17,tang+18}. Within the standard $\lesssim 10$~keV band accessible to Chandra, these effects are therefore expected to manifest primarily as modest spectral hardening or increased spectral complexity, rather than as distinct high-energy components. As a result, CCD-resolution spectra can be difficult to distinguish from those of single-SMBH AGN affected by absorption or disk reflection.  Recent theoretical work further emphasizes that such degeneracies are intrinsic to CCD-resolution X-ray data, even when the underlying emission arises from binary SMBH accretion flows. In particular, \citet{malewicz+25} showed that unequal-mass binaries with multiple mini-disks can produce composite reflection spectra spanning a wide range of ionization states, yet still be acceptably fit by standard single-AGN models in the low-count regime. Although their calculations focused on binaries at separations of $a = 100\,R_g$, this lies in a regime where relativistic effects remain comparatively weak, allowing their qualitative conclusions regarding spectral degeneracies to extend to wider-separation candidate SMBH binaries.

\subsection{Time-Domain and Spectroscopic Signatures of Binary SMBHs in AGN}
\label{intro:time_domain}
In contrast to the model-driven X-ray signatures discussed above, most observational searches for binary SMBHs have relied on temporal variability and spectroscopic diagnostics.  Beyond X-ray spectral diagnostics, an alternative and widely discussed signature of binary SMBHs is periodic or quasi-periodic variability in quasar light curves, where the periodicity may arise from orbital modulation of the accretion flow or relativistic Doppler boosting \citep[e.g.,][]{graham+15,oneill+21,charisi+22}.  However, several alternative mechanisms can also produce periodic or temporarily quasi-periodic behavior, including accretion-disk instabilities \citep{king+07}, warped disks \citep{hopkins+10}, jet precession \citep[e.g.,][]{caproni+04,rieger+04}, or stochastic red-noise variability \citep[e.g.,][]{vaughan+16}.  

A complementary observational signature is found in quasars whose broad emission lines are systematically displaced from the host-galaxy systemic velocity \citep[e.g.,][]{bogdanovic+09,tsalmantza+11,eracleous+12,ju+13,shen+13,runnoe+17}.  These ``velocity-offset'' broad lines can naturally arise if the broad line region (BLR) is gravitationally bound to one SMBH in a binary and the binary orbital motion leads to the periodic Doppler shifting of its broad emission lines \citep[][]{runnoe+17}.  If interpreted as orbital motion in a bound SMBH binary, the observed velocity offsets of
$\sim10^{3}\,\mathrm{km\,s^{-1}}$ imply characteristic orbital separations of order
$a \sim 10^{3}$--$10^{4}\,R_g$ for black hole masses typical of luminous quasars
($M_{\bullet}\sim10^{8}\,M_{\odot}$).  
Such separations are also consistent with the multi-year persistence and relatively slow
centroid drifts of the broad-line velocity offsets observed in spectroscopic monitoring
campaigns \citep[e.g.,][]{runnoe+17,mohammed+25}.  

However, alternative physical scenarios can also produce broad-line velocity offsets.  A SMBH that has received a velocity kick following binary SMBH coalescence may retain its BLR as it moves through the host galaxy's gravitational potential \citep[e.g.,][]{loeb07,komossa+12,chiaberge+17,chiaberge+18}. 
Large recoil velocities can also arise from dynamical three-body interactions between a bound SMBH binary and a third SMBH interloper, with speeds that can exceed $1,000~\mathrm{km~s^{-1}}$ \citep{hoffman_loeb_06}.  Alternatively, the velocity offsets may result from outflows or other feedback processes that impart momentum to the BLR gas, for example via a relativistic jet, an accretion-disk wind, or even a supernova explosion \citep[e.g.,][]{yong+17,breiding+21,rusakov+23,matthews+23}.  The offsets may also arise from a perturbed or disk-like BLR, in which asymmetric illumination or non-axisymmetric emissivity within an accretion disk or disk–wind structure produces displaced broad-line centroids without requiring bulk motion of the SMBH \citep[e.g.,][]{gezari+07,tang+09,gaskell10,popovic12}.

\subsection{X-ray Observations of Velocity-Offset Quasars as Binary SMBH Candidates}

Motivated by the theoretical predictions and observational ambiguities outlined above, we have carried out a study using new Chandra observations of five nearby ($0.1<z<0.2$) quasars drawn from the velocity-offset quasar sample of \citet{eracleous+12}, supplemented with archival data for an additional eleven sources from the same sample.  The \citet{eracleous+12} sample comprises 88 quasars from the Sloan Digital Sky Survey (SDSS) exhibiting $\gtrsim$1000~km~s$^{-1}$ velocity shifts of their broad Balmer $\mathrm{H\beta}$ emission lines relative to the narrow [$\mathrm{O\,III}$]~$\lambda5007$ lines that define the host-galaxy rest frame.  Previous optical spectroscopic monitoring over multi-year baselines revealed systematic variations in the velocity centroids of several objects within the broader parent sample, consistent with orbital motion \citep[][]{runnoe+17}, while very long baseline interferometry (VLBI) observations of 34 sources provided no definitive evidence for or against the binary-SMBH interpretation \citep{breiding+21}.  This X-ray study of the \citet{eracleous+12} velocity-offset quasar sample aims to investigate the nature of these systems by probing their high-energy emission and constraining the competing physical models proposed to explain their broad-line velocity shifts.

Throughout this paper we adopt a flat $\Lambda$CDM cosmology with $H_{0}=67.74~\mathrm{km\,s^{-1}\,Mpc^{-1}}$, $\Omega_{\Lambda}=0.69$, and $\Omega_{m}=0.31$ \citep{planck+16}.




\begin{figure}
	\centering
	\includegraphics[width=\linewidth]{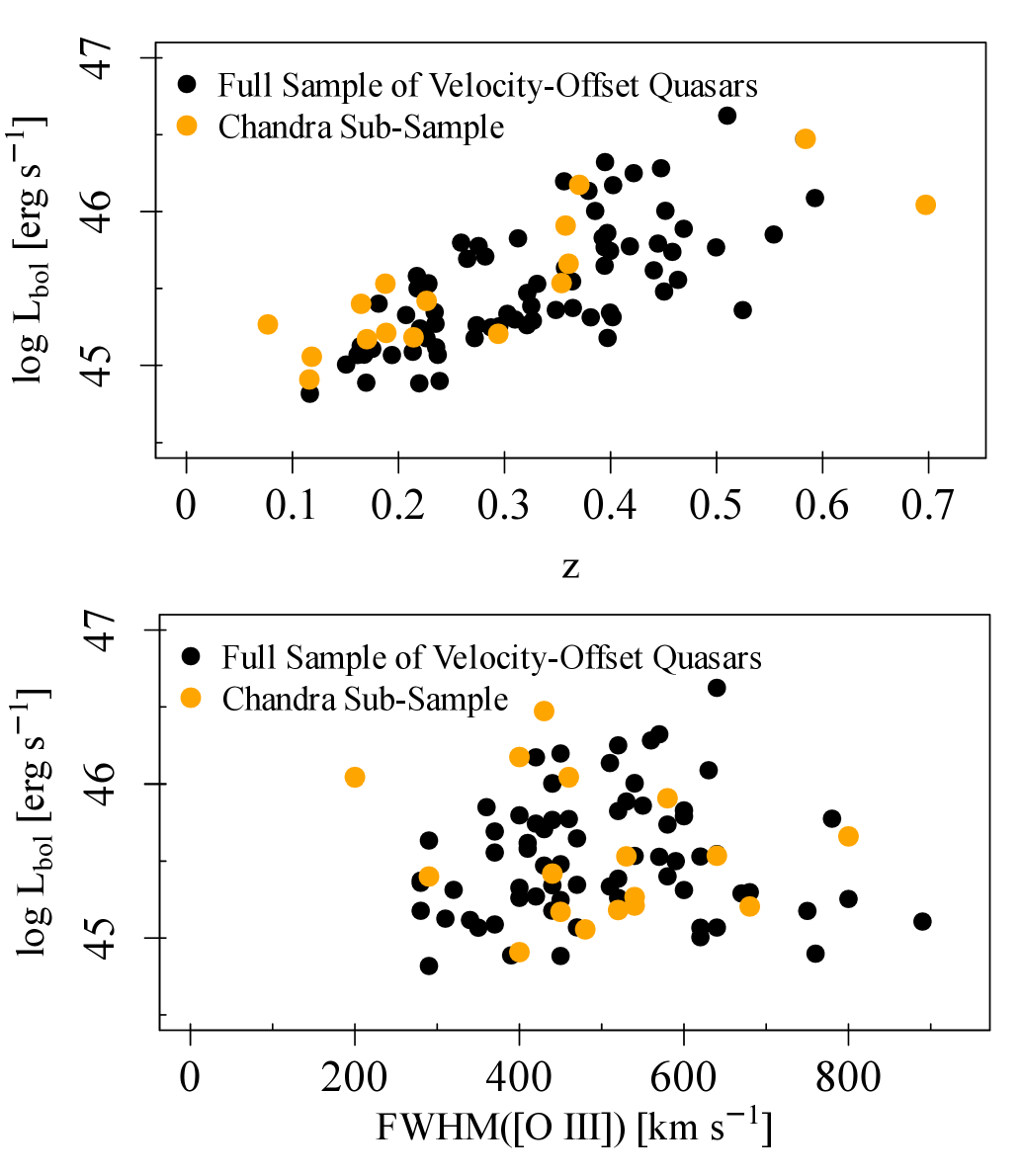}
	\caption{
		\textit{Top:} Bolometric luminosity as a function of redshift for the full \citet{eracleous+12} sample of velocity-offset quasars (black points) and the Chandra-observed sub-sample analyzed in this work (orange points).  Bolometric luminosities are taken from \citet{shen+11}.
		\textit{Bottom:} Bolometric luminosity as a function of the narrow-line FWHM([O\,\textsc{iii}]~$\lambda5007$) for the same two samples, with matching symbols and colors.  Narrow-line widths are taken from Table~3 of \citet{eracleous+12}.
		Together, these panels show that the Chandra sub-sample spans similar ranges in luminosity and narrow-line kinematics as the parent sample.
	}
	\label{fig:parent_comp}
\end{figure}

\section{Chandra Observations} \label{sec:obs}

The new observations associated with this work were executed as Chandra guest observer (GO) time during Cycle 25 (P.I.  P.  Breiding), and are summarized in Table~\ref{table:cxo_obs}.  We summarize the archival Chandra observations used in this work in Table~\ref{table:cxo_arch}.  In both Tables~\ref{table:cxo_obs} and \ref{table:cxo_arch} we give the full SDSS names of the quasars, which reflect their J2000 equatorial coordinates. Throughout the rest of the paper we truncate the source names to the first six digits of Right Ascension for brevity.  Our Chandra observations summarized in Table~\ref{table:cxo_obs} were designed to achieve $\sim$20~ks exposures for the subset of the parent sample expected to be brightest in X-rays.  The targets were selected based on their [O\,\textsc{iii}]\,$\lambda5007$ line luminosities, which are known to correlate with intrinsic X-ray luminosity in AGN \citep[e.g.,][]{shen+06}.
These exposure times were chosen to obtain high signal-to-noise spectra, minimize uncertainties in derived spectral parameters, and provide a sufficiently long baseline for the light curve analysis described in Section~\ref{subsec:light_curve_analysis}.  In practice, a small number of observations were executed with shorter individual exposures or split into multiple segments due to scheduling considerations, rather than differing scientific goals, and in all cases the achieved exposure times were sufficient to meet the primary objectives of the program. 

\begin{table*}
	\centering
	\caption{New Chandra Observations} 
	\label{table:cxo_obs}
	\begin{threeparttable}
		\begin{tabular}{llcllllr} 
			\hline
			SDSS&Abbrev.&z&Obs.&Obs.&Instrument&Data&Exposure\\
			Name&Name&&ID&Date&&Mode&(ks)\\
			\hline
			SDSS~J015530.02$-$085704.0&J015530&0.165&28298&10/19/2024&ACIS~S&VFAINT&20.8\\
			SDSS~J093844.46+005715.7&J093844&0.170&28299$^{\dagger}$&03/09/2024&ACIS~S&VFAINT&11.2\\
			SDSS~J093844.46+005715.7&J093844&0.170&29313$^{\dagger}$&03/10/2024&ACIS~S&VFAINT&11.2\\
			SDSS~J113651.66+445016.4&J113651&0.116&28297&04/20/2024&ACIS~S&VFAINT&20.3\\
			SDSS~J125142.28+240435.3&J125142&0.188&28296$^{\dagger\dagger}$&03/31/2024 &ACIS~S&VFAINT&11.4\\
			SDSS~J125142.28+240435.3&J125142&0.188&29349$^{\dagger\dagger}$&03/31/2024 &ACIS~S&VFAINT&11.4\\
			SDSS~J130534.49+181932.9&J130534&0.118&28295&04/07/2024&ACIS~S&VFAINT&18.8\\
			\hline
		\end{tabular}
		\begin{tablenotes}	
			\item[] Observation dates are given in the format MM/DD/YYYY.
			\item[$\dagger$] These observations were combined.
			\item[$\dagger\dagger$] These observations were combined.
		\end{tablenotes}
	\end{threeparttable}
\end{table*}

\begin{table*}
	\centering
	\caption{Archival Chandra Observations} 
	\label{table:cxo_arch}
	\begin{threeparttable}
		\begin{tabular}{llcllllr} 
			\hline
			SDSS&Abbrev.&z&Obs.&Obs.&Instrument&Data&Exposure\\
			Name&Name&&ID&Date&&Mode&(ks)\\
			\hline
			SDSS~J020011.53$-$093126.2&J020011&0.360&15036$^{\dagger}$&11/05/2013 &ACIS~S&VFAINT&9.7\\
			SDSS~J020011.53$-$093126.2&J020011&0.360&15577$^{\dagger}$&11/05/2013 &ACIS~S&VFAINT&19.3\\
			SDSS~J092712.65+294344.0&J092712&0.697&10721&02/23/2009&ACIS~S&FAINT&26.7\\
			SDSS~J095036.75+512838.1&J095036&0.214&23328$^{\dagger\dagger}$&11/05/2020&ACIS~S&VFAINT&5.0\\
			SDSS~J095036.75+512838.1&J095036&0.214&23821$^{\dagger\dagger}$&01/15/2021&ACIS~S&VFAINT&13.0\\
			SDSS~J102839.11+450009.3&J102839&0.584&17097&10/18/2015&ACIS~S&VFAINT&4.4\\
			SDSS~J110556.19+031243.2&J110556&0.354&9196&01/24/2008&ACIS~S&FAINT&4.1\\
			SDSS~J121113.97+464711.9&J121113&0.294&20770&08/05/2018&ACIS~S&VFAINT&14.9\\
			SDSS~J124551.03+032128.4&J124551&0.227&12288&12/05/2010&ACIS~I&VFAINT&90.8\\
			SDSS~J140251.20+263117.5&J140251&0.188&2166&07/04/2001&ACIS~I&VFAINT&1.7\\
			SDSS~J140700.40+282714.6&J140700&0.077&16071&09/04/2014&ACIS~S&VFAINT&34.6\\
			SDSS~J151443.07+365050.4&J151443&0.371&3988&10/05/2003&ACIS~I&VFAINT&41.6\\
			SDSS~J165118.62+400124.8&J165118&0.358&15872&05/19/2014&ACIS~S&VFAINT&19.8\\
			\hline
		\end{tabular}
		\begin{tablenotes}	
			\item[] Observation dates are given in the format MM/DD/YYYY.
			\item[$\dagger$] These observations were combined.
			\item[$\dagger\dagger$] These observations were combined.
		\end{tablenotes}
	\end{threeparttable}
\end{table*}

In order to increase the sample size, and thus robustness of our study, we searched the Chandra data archive for X-ray observations of the other quasars from our parent sample.  To do this, we performed a cone search for archival Chandra observations which had a pointing within 30\arcmin of our quasar positions.  We then checked whether our quasars fell onto any ACIS chips for the observations matching the above criteria, leading to the list of observations summarized in Table~\ref{table:cxo_arch}.  Most of the archival observations listed in Table~\ref{table:cxo_arch} were designed with the quasars analyzed in this work as their principal science targets.  However, J121113, J124551, J140007, and J165118 were observed serendipitously because they were in the fields of different primary science targets.  In Figure~\ref{fig:parent_comp}, we compare the redshift–bolometric luminosity and bolometric luminosity–narrow-line FWHM distributions of our Chandra sub-sample to the full set of 88 velocity-offset quasars from \citet{eracleous+12}.  This comparison shows that the quasars targeted for X-ray follow-up span similar ranges in luminosity, redshift, and narrow-line kinematics as the parent population.  Although our analysis focuses on only 16 of the 88 objects ($\sim18\%$ of the parent sample), the Chandra targets appear broadly representative of the full velocity-offset quasar sample in these global properties.     

\section{Data Reduction \& Analysis} \label{sec:analysis}

All Chandra data were reduced with the Chandra Interactive Analysis of Observations (CIAO) software package \citep{ciao}, version 4.17.  The data were first processed with the \texttt{chandra\_repro} script, applying the most up-to-date calibrations at the time of analysis (CALDB 4.12.1).  

\subsection{X-ray Spectral Analysis}
\label{subsec:spectral_analysis}

Spectra were extracted using the \texttt{specextract} CIAO task using circular apertures of radius 4\arcsec, with background spectra taken from surrounding annuli spanning radii of 9\arcsec to 46\arcsec.  Within \texttt{specextract} we set weight=no to generate point-source ancillary response files (ARFs) and correctpsf=yes to apply the encircled-energy correction appropriate for source extraction regions.  For sources with multiple Chandra observations, we combined the spectra before fitting using the CIAO task \texttt{combine\_spectra}.

Spectral fits were performed ungrouped (i.e., not binned) using the C statistic \citep{Cash1979}, which is based on the Poisson likelihood function which is appropriate for X-ray counts data.  The C statistic provides unbiased parameter estimation, particularly in the low-counts regime where the Gaussian assumptions underlying fits relying on $\chi^2$ statistics break down \citep{Cash1979,Humphrey2009,Kaastra2017}.   

All spectra were fit in the observer-frame $0.5$--$7$~keV band using a simple redshifted power law model.  In each spectral fit, we also included a component accounting for absorption by the Milky Way Galactic neutral hydrogen column along the line of sight\footnote{Galactic hydrogen column densities were obtained from the NASA HEASARC web tool: \url{https://heasarc.gsfc.nasa.gov/cgi-bin/Tools/w3nh/w3nh.pl}}.  We also jointly fit the background with a simple power law model, with a Milky Way hydrogen column absorption component.  For some sources we also included an intrinsic hydrogen column absorption component.   

To assess whether an intrinsic absorption component was statistically required for each source, we performed Monte Carlo simulations.  For each quasar spectrum we tested for intrinsic absorption by comparing two nested models:

\begin{itemize}
	\item[$H_{0}$:] The null hypothesis --  A baseline model consisting of a redshifted power law attenuated only by the Milky Way's absorbing neutral hydrogen column.
	\item[$H_{1}$:] The alternative hypothesis -- A redshifted power-law model with a Milky Way absorption component and an additional intrinsic absorber at the source redshift.
\end{itemize}

We then computed the likelihood ratio test statistic as the difference in C statistics between the models, $\Delta C = C(H_{0}) - C(H_{1})$, where this equality follows from the definition of the C statistic in terms of the log likelihood: $C=-2~\ln~L$.  To evaluate the significance of the observed test statistic, we generated 1000 simulated spectra under the null hypothesis using the Sherpa task \texttt{fake\_pha}.  The Monte Carlo $p$-value was then taken as the fraction of simulated test statistics exceeding the observed $\Delta C$, and intrinsic absorption was adopted only for sources with $p < 0.05$.

\subsubsection{Fe K$\alpha$ Emission Line Searches}
\label{subsubsec:line_detection}

For each quasar, we searched for Fe~K$\alpha$ emission by modeling putative lines with a Gaussian profile whose centroid energy was allowed to vary between 6.25~keV and 6.6~keV, and whose standard deviation ($\sigma$) was permitted to span 0.01–0.40~keV.  
To reduce false positives in low-count spectra, we required a minimum equivalent width of $\mathrm{EW} > 40$~eV when evaluating potential line candidates. This threshold lies near the lower end of narrow Fe\,K$\alpha$ equivalent widths typically measured in unobscured Type~1 AGN \citep[e.g.,][]{yaqoob+04,shu+10,ricci+13,ricci+14}.  For all candidate lines passing these preliminary criteria, we performed a Monte Carlo significance test following the same approach used in Section~\ref{subsec:spectral_analysis} for evaluating intrinsic absorption.  In each simulation, synthetic spectra were drawn from the best-fitting baseline continuum model (the null hypothesis), refit in the same manner as the real data, and the distribution of $\Delta C$ improvements from adding a Gaussian line was recorded.  The line was considered detected if fewer than 5\% of the simulated continuum-only spectra produced a value of $\Delta C$ equal to or larger than that observed in the real data (i.e., $p<0.05$).  For all quasars where the emission-line model was preferred according to this Monte Carlo test, we included the Gaussian component in all subsequent spectral modeling and parameter estimation.  

\subsection{Light Curve Analysis}
\label{subsec:light_curve_analysis} 

We extracted background-subtracted light curves for each source in the rest-frame $2$--$10$~keV band using the CIAO tool \texttt{dmextract}, with source and background regions identical to those used for spectral extraction.  We generated light curves with a series of trial time bin sizes, beginning at $250$~s in the source rest frame and increasing in steps of $50$~s, until a binning was found that provided at least $20$ counts per bin.  This threshold ensures that the Gaussian distribution provides a reasonable approximation to the Poisson uncertainties in each time bin \citep[e.g.,][]{gehrels86}.  If this procedure produced fewer than $20$ total bins in the resulting light curve, we did not proceed with the variability analysis\footnote{A sufficient number of bins is required to obtain a reliable estimate of the normalized excess variance and its uncertainty \citep[e.g.,][]{vaughan+03}.}.   Time bins with fractional exposures below 0.35 were discarded.  For sources with multiple Chandra observations, we combined the background-subtracted light curves with the CIAO tool \texttt{dmmerge} to analyze them as one contiguous light curve. 

For each light curve we computed the normalized excess variance following \cite{Nandra+97}.  This quantity measures intrinsic source flux variability in excess of that expected from random measurement errors.  It is defined as:

\begin{equation}
	\sigma_{\mathrm{NXS}}^{2} \;=\; 
	\frac{1}{N \mu^{2}} \sum_{i=1}^{N} \left[ (x_i - \mu)^{2} - \sigma_i^{2} \right],
\end{equation}
\\
where $N$ is the number of light curve time bins, 
$x_i$ is the $i$-th count rate, 
$\mu$ is the mean count rate, 
and $\sigma_i$ is the measurement error for $x_i$.

We adopt the uncertainty formulation for the normalized excess variance from \cite{vaughan+03}, 
\begin{equation}
	\mathrm{err}(\sigma_{\mathrm{NXS}}^{2}) \;=\;
	\sqrt{ \left( \sqrt{\tfrac{2}{N}} \, \frac{\overline{\sigma^{2}}}{\mu^{2}} \right)^{2}
		+ \left( \sqrt{\tfrac{\overline{\sigma^{2}}}{N}} \, \frac{2 F_{\mathrm{var}}}{\mu} \right)^{2} },
\end{equation}
where $\overline{\sigma^{2}} = \tfrac{1}{N}\sum_{i=1}^{N}\sigma_i^{2}$ is the mean square of the measurement uncertainties 
and $F_{\mathrm{var}} = \sqrt{\sigma_{\mathrm{NXS}}^{2}}$ is the fractional variability amplitude.  This prescription includes both the contribution from measurement errors and an additional term that accounts for the intrinsic stochastic variance of the source flux.  We define variability non-detections as cases where the $\mathrm{\sigma^{2}_{NXS}}$ $1\sigma$ confidence interval includes 0, in which case we report 95\% upper limits on $\mathrm{\sigma^{2}_{NXS}}$.

To enable comparison with the $\sigma_{\mathrm{NXS}}^{2}$--$M_{\mathrm{\bullet}}$ scalings of \cite{ponti+12}, we corrected our normalized excess variance measurements to a common reference frequency band.  Assuming a single power-law power spectral density (PSD) of the form $P(f)\propto f^{-\alpha}$ (where $f$ is frequency, $P$ is the power, and $\alpha$ is the PSD slope), the normalized excess variance can be shown to follow the proportionality,
\begin{equation}
	\sigma^{2}_{NXS} \;\propto\; \int_{f_{\mathrm{low}}}^{f_{\mathrm{high}}} f^{-\alpha}\, df,
\end{equation}
with $f_{\mathrm{low}} = 1/T$, $f_{\mathrm{high}} = 1/(2\Delta t)$, $T$ being the observation duration, and $\Delta t$ being the time bin size,   \citep[e.g.,][]{vanderklis89, vaughan+03}.  Since our sampled frequency bands in general differ from the frequency bands sampled by \cite{ponti+12}, where the authors use 250~s time bins and observations of either 10~ks, 20~ks, or 40~ks, we must correct our measurements of normalized excess variance for the change in our sampled frequency band to that sampled by \cite{ponti+12}.  We determine the correction factors applied to the measurement of $\sigma_{\mathrm{NXS}}^{2}$ in each source as the ratio of the integrated PSD variance in the reference frequency band, corresponding to the particular \cite{ponti+12} scaling we are using, to that in the observed frequency band:
\begin{equation}
	\label{eq:excess_var_correction}
	A \;=\; \frac{\int_{1/T_{\mathrm{ref}}}^{1/(2\Delta t_{\mathrm{ref}})} f^{-\alpha}\, df}
	{\int_{1/T_{\mathrm{obs}}}^{1/(2\Delta t_{\mathrm{obs}})} f^{-\alpha}\, df}.
\end{equation}


We adopt a PSD slope of $\alpha = 2$, which is 
consistent with the high-frequency PSD slope above the break frequency\footnote{The break frequency corresponds to the frequency where the PSD slope changes (usually assumed to change from $\alpha = 1$ to $\alpha = 2$), as determined by the viscous timescale of the inner accretion flow \citep{lyubarskii+97}.}, as found in previous AGN studies \citep[][]{markowitz+03,mchardy+04,gonzalez+12}.  The PSD break frequency can be derived from the $M_{\rm BH}$--$L_{\rm bol}$ AGN-only relation of \cite{mchardy+06} as:
\begin{equation}
	\label{break_eqn}
	f_{\rm b} \;\simeq\; 3.75\times10^{-3}\ {\rm Hz}\;
	\left(\frac{M_{\rm BH}}{10^{6}M_\odot}\right)^{-1.27}
	\left(\lambda\right)^{0.90},
\end{equation}
where $\lambda=L_{\rm bol}/L_{\rm Edd}$ is the Eddington ratio, $L_{bol}$ is the bolometric luminosity, and $L_{Edd}$ is the Eddington luminosity.  Since the minimum sampled frequency of our variability analysis is given by $f_{\min}=1/T$, and a representative rest-frame light curve length used in our analysis is 20~ks, the minimum sampled frequency for estimating $\sigma_{\mathrm{NXS}}^{2}$ is $\sim5.0\times10^{-5}$~Hz.  Assuming $\mathrm{log(M_{\bullet})=8}$ and $\mathrm{\lambda = 0.1}$, both representative for our quasars as shown in Section~\ref{results}, we derive a break frequency of $f_{b}\sim1.4\times10^{-6}$~Hz from Equation~\ref{break_eqn}, implying that our sampled frequency range is well above the break frequency for our variability analyses.  

We scaled the normalized excess variance by the factor $A$ given in Equation~\ref{eq:excess_var_correction} and scaled its uncertainty by the same factor, such that $\sigma_{\mathrm{NXS,\ corr}}^{2} = A\,\sigma_{\mathrm{NXS,\ obs}}^{2}$.  This procedure places our measurements of $\sigma_{\mathrm{NXS}}^{2}$ on the same frequency range used by \cite{ponti+12}, enabling direct comparison across sources with different observation lengths and time binning.

\section{Results}
\label{results} 

\subsection{Spectral Fitting}
\label{subsec:fitting_results}  

\begin{figure*}
	\includegraphics[width=\textwidth]{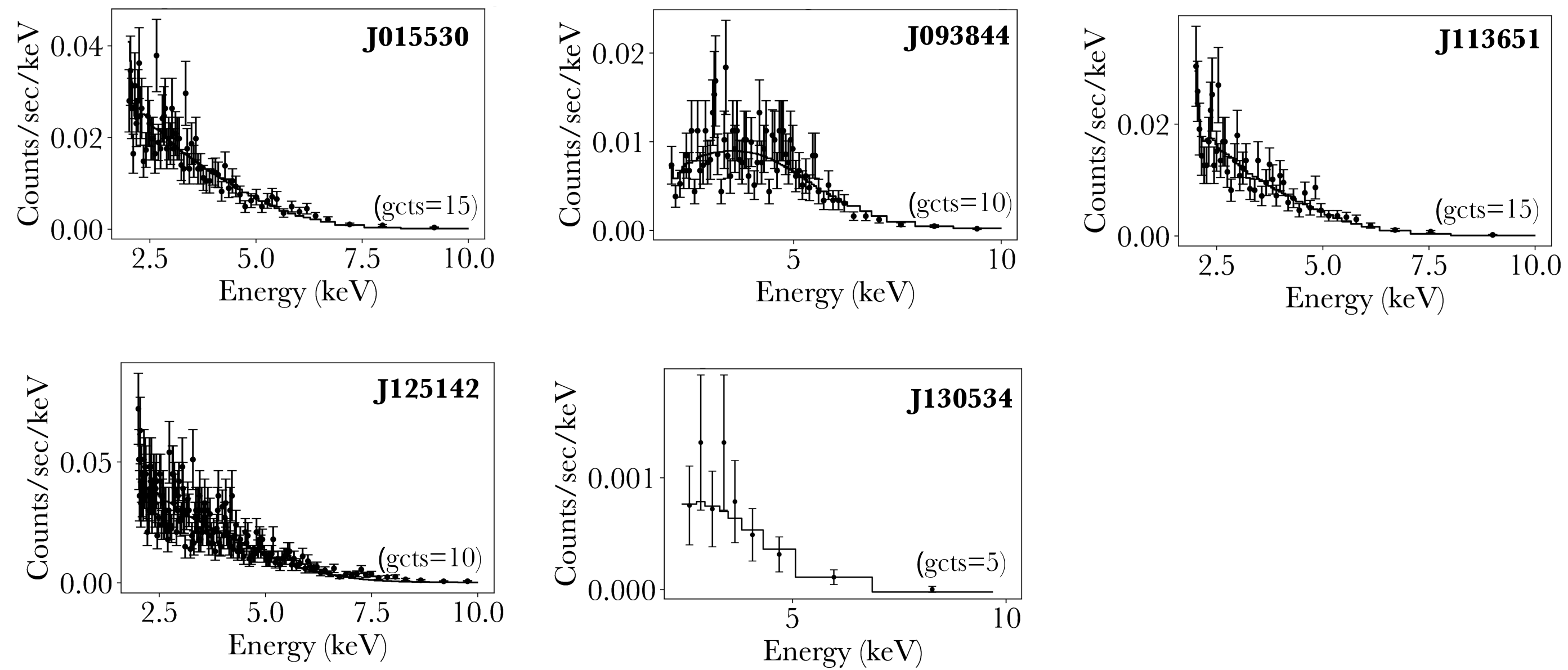}
	\caption{Spectral fits in the 2$-$10~keV observer frame for the Chandra GO observations detailed in Table~\ref{table:cxo_obs}.  The model is shown as a solid black curve, with data points plotted using 1$\sigma$ error bars.  Spectra are grouped for visualization (the grouping is indicated by ``gcts'' in each panel), while all fits are performed on unbinned data.  Source names are given in the upper right corner of each spectrum.}
	\label{fig:spectral_fits}
\end{figure*} 

Figure~\ref{fig:spectral_fits} shows the spectral fits for the five quasars listed in Table~\ref{table:cxo_obs}.  Table~\ref{table:spectral_fitting} summarizes the best-fit spectral parameters and the observational quantities derived from our modeling.  In Table~\ref{table:spectral_fitting} we list the intrinsic hydrogen column density, $N_{H}$, for sources with an intrinsic absorption component in the spectral model, the X-ray photon index, $\Gamma$, the rest-frame $2-10$~keV broadband flux, $\mathrm{F_{2-10~keV}}$, the rest-frame 2~keV flux density, $\mathrm{f_{2~keV}}$, and the number of $0.5-7$~keV counts used in the fit.  Flux density and flux uncertainties were estimated with the Sherpa task \texttt{sample\_energy\_flux}.  Uncertainties were derived via Monte Carlo sampling, drawing 1000 realizations of the model parameters from a multivariate normal distribution centered on the best-fit values.  Reported errors are $1\sigma$ uncertainties centered around the median values.   
For two low-count sources with intrinsic absorption (J121113 and J130534), we instead report the 95\% confidence intervals, as the sampled distributions were highly asymmetric about the median values.

\begin{table*}
	\centering
	\caption{Results of X-ray Spectral Analysis} 
	\label{table:spectral_fitting}
	\begin{threeparttable}
		\begin{tabular}{lcccccc} 
			\hline
			Source&$\mathrm{N_{H}}$&$\mathrm{\Gamma_{0.5-7~keV}^{\dagger\dagger}}$&$\mathrm{F^{\dagger}_{2-10~keV}}$&$\mathrm{f^{\dagger}_{2~keV}}$&Counts ($0.5-7$~keV)$^{\dagger\dagger}$&$\mathrm{\alpha_{ox}}$\\
			Name&$\mathrm{(10^{22}~cm^{-2})}$&&($\mathrm{1\times10^{-14}~erg~s^{-1}cm^{-2}}$)&(nJy)&&\\
			\hline
			J015530&...&$1.78^{+0.04}_{-0.04}$&136$\pm11$&$168.0\pm9.3$&2283&$1.37\pm0.11$\\
			J020011&...&$1.75^{+0.04}_{-0.04}$&$38.6\pm2.7$&$52.5\pm2.9$&1899&$1.41\pm0.11$\\
			J092712&...&$0.86^{+0.21}_{-0.21}$&$2.06\pm1.11$&$1.6\pm0.5$&66&$1.92\pm0.12$\\
			J093844&2.0&$0.51^{+0.17}_{-0.17}$&$114\pm59$&$19.0\pm7.5$&820&$1.32\pm0.13$\\
			J095036&...&$1.93^{+0.14}_{-0.14}$&$10.2\pm2.8$&$15.0\pm2.8$&196&$1.46\pm0.11$\\
			J102839&...&$2.11^{+0.23}_{-0.22}$&$8.69\pm4.23$&$18.9\pm5.2$&73&$1.68\pm0.12$\\
			J110556&1.1&$0.91^{+0.26}_{-0.23}$&$57\pm35$&$24.9\pm4.2$&205&$1.13\pm0.11$\\
			J113651&...&$1.90^{+0.05}_{-0.05}$&$89.4\pm8.8$&$117.1\pm6.7$&1601&$1.26\pm0.11$\\
			J121113&29.3&$3.07^{+3.04}_{-2.08}$&$^{\dagger\dagger\dagger}$$18.7^{+110}_{-18.7}$&$^{\dagger\dagger\dagger}$$0.35^{+9.04}_{-0.34}$&64&$^{\dagger\dagger\dagger}$$2.15^{+0.83}_{-0.58}$\\
			J124551&...&$0.11^{+0.05}_{-0.05}$&$43.7\pm4.2$&$10.4\pm0.7$&1547&$1.74\pm0.11$\\
			J125142&...&$1.58^{+0.03}_{-0.03}$&$228\pm13$&$239.0\pm10.2$&3828&$1.18\pm0.11$\\
			J130534&4.1&$2.35^{+0.91}_{-0.82}$&$^{\dagger\dagger\dagger}$$2.57^{+480}_{-2.57}$&$^{\dagger\dagger\dagger}$$0.22^{+38.72}_{-0.22}$&47&$^{\dagger\dagger\dagger}$$2.09^{+0.70}_{-0.36}$\\
			J140251&...&$1.91^{+0.10}_{-0.10}$&$121\pm21$&$169.9\pm17.0$&363&$1.32\pm0.11$\\
			J140700&...&$1.19^{+0.07}_{-0.07}$&$28.7\pm3.3$&$15.1\pm1.4$&711&$1.61\pm0.11$\\
			J151443&...&$1.35^{+0.02}_{-0.02}$&$180\pm8$&$178.0\pm4.6$&7654&$1.39\pm0.11$\\
			J165118&0.8&$1.00^{+0.15}_{-0.14}$&$56.6\pm21.7$&$29.1\pm7.7$&615&$1.60\pm0.12$\\
			\hline
		\end{tabular}
		\begin{tablenotes}	
			\item[] The reported $N_{\rm H}$ values represent the intrinsic absorbing column in excess of the Galactic foreground column, which is fixed to the Milky Way value along the line of sight.
			\item[$\dagger$] These quantities are given in the rest-frame $2-10$~keV band.
			\item[$\dagger\dagger$] These quantities are given in the observer-frame $0.5-7$~keV band.
			\item[$\dagger\dagger\dagger$] These error bars represent 95\% confidence intervals.  All other error bars represent 68\% confidence intervals.
		\end{tablenotes}
	\end{threeparttable}
\end{table*}

\subsubsection{Fe K$\alpha$ Emission Line Detections}
\label{subsubsec:line_results}

\begin{table}
	\caption{Fe~K$\alpha$ Emission Line Detections}
	\label{table:line_detections}
	\begin{threeparttable}
		\begin{tabular*}{\linewidth}{@{\extracolsep{\fill}}lcrcr}
			\hline
			Source&EW&FWHM&Line Energy&p-val\\
			Name&(eV)&(keV)&(keV)&\\
			\hline
			J020011&$271^{+120}_{-114}$&$<0.66$&$6.28^{+0.08}_{-0.08}$&0.007\\
			J092712&$464^{+324}_{-261}$&...&$6.47^{+0.09}_{-0.09}$&0.001\\
			J113651&$128^{+83}_{-72}$&...&$6.50^{+0.06}_{-0.06}$&0.02\\
			J121113&$250^{+170}_{-134}$&...&$6.34^{+0.05}_{-0.06}$&0.039\\
			J125142&$110^{+56}_{-55}$&...&$6.56^{+0.07}_{-0.06}$&0.009\\
			J140700&$2114^{+727}_{-635}$&$0.65^{+0.18}_{-0.15}$&$6.51^{+0.07}_{-0.07}$&0.001\\
			\hline
		\end{tabular*}
		\begin{tablenotes}[flushleft]
			\item[] Line energies are given in the rest frame.  For lines with best-fit FWHM values $<$150~eV, we do not report a FWHM.  EW uncertainties were estimated by sampling the covariance matrix of the best-fit spectral model parameters and recomputing the EW for each realization; we report the median EW and the central 68\% confidence interval.
		\end{tablenotes}
	\end{threeparttable}
\end{table}



\begin{figure*}
	\includegraphics[width=\textwidth]{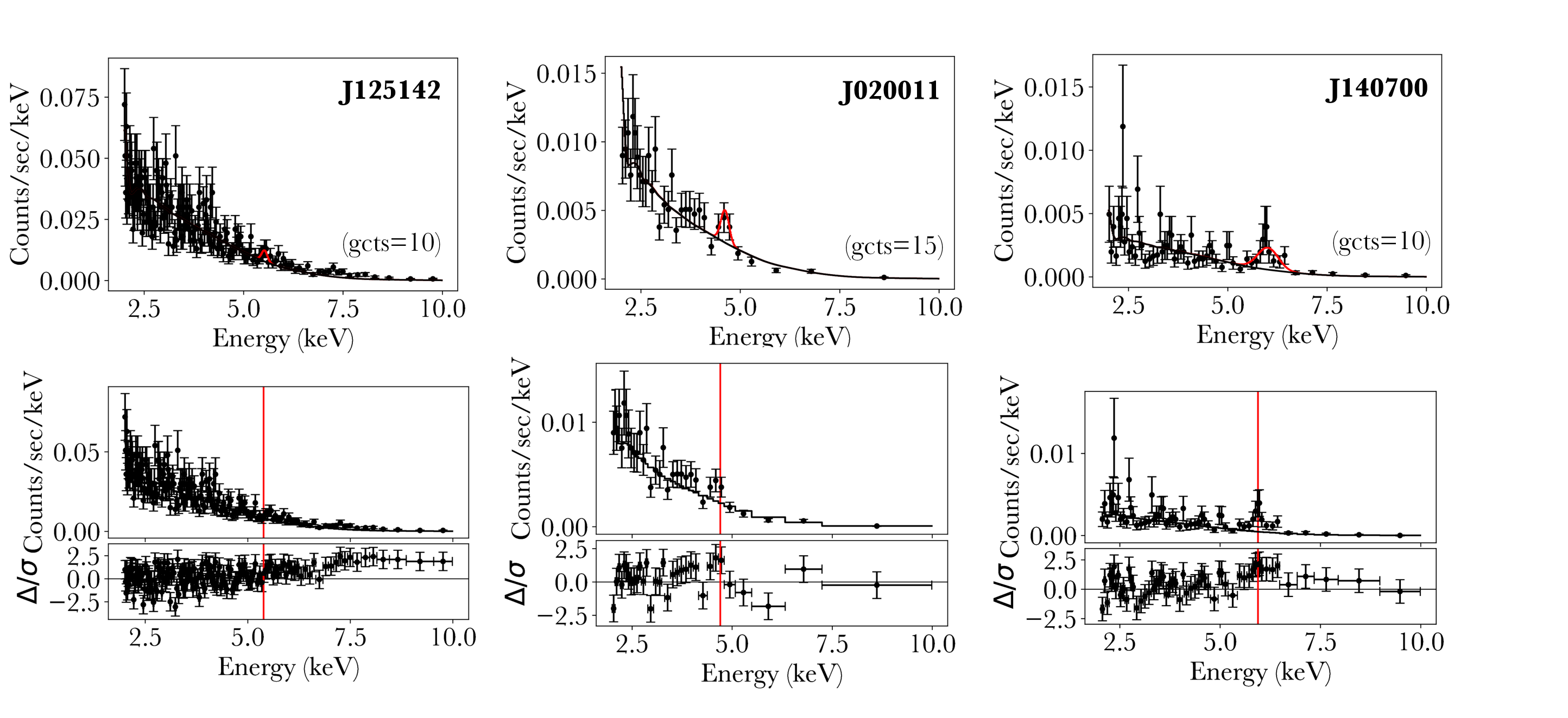}
	\caption{\textit{Top row}: Observer-frame $2-10$~keV spectra of three representative sources exhibiting Fe~K$\alpha$ emission.  These examples illustrate, respectively, an unresolved line (left panel), a potentially moderately resolved line (middle panel), and a fully resolved line (right panel).  Solid black curves denote model fits without an emission-line component, while red curves include an Fe~K$\alpha$ line model component.  Source names are shown in the upper right; ``gcts'' gives the minimum number of photons per energy bin (spectra are grouped only for visualization, as all fitting is performed on unbinned data).  Source names and gcts values also apply to the corresponding plots in the bottom row.  
		\textit{Bottom row}: Solid black curves show model fits without an emission-line component, with the corresponding residuals plotted below.  Vertical red lines mark the observed-frame energies corresponding to 6.4~keV in the rest frame.}
	\label{fig:line_fits}
\end{figure*}

We detect Fe~K$\alpha$ emission lines in 6 of the 16 quasars in our sample.  The line-detection properties are summarized in Table~\ref{table:line_detections}, where we list the fitted spectral parameters and the associated detection significance.  Four of the six FeK$\alpha$ lines are clearly unresolved, with the best-fit models having FWHM values $<150$~eV, consistent with the instrumental resolution of Chandra’s ACIS~S detector\footnote{The on-axis ACIS-S FWHM energy resolution near 6~keV is $\sim150$~eV (see the \textit{Chandra Proposers’ Observatory Guide}: \url{https://cxc.harvard.edu/proposer/POG/}).}.  For J020011, the best-fit model prefers a moderately broadened line with ${\rm FWHM} \simeq 0.27$~keV (corresponding to a velocity width of $\sim$12{,}900km~s$^{-1}$, or $\sim$0.04\,$c$); however, the 1$\sigma$ uncertainty on the line width includes solutions consistent with an unresolved line.  As a result, the line broadening in J020011 is not statistically significant, and we conservatively report a 95\% upper limit of ${\rm FWHM} < 0.66$~keV.  By contrast, J140700 exhibits a clearly broadened FeK$\alpha$ line with a FWHM of $0.64^{+0.18}_{-0.15}$keV, implying a velocity width of $\sim$29{,}400km~s$^{-1}$ ($\sim$0.1\,$c$).  The implied projected velocities for J020011 and J140700 ($v \sim 0.04$–$0.10\,c$) correspond to orbital speeds at radii of $\sim10^2$–$10^3,R_g$, with the upper end overlapping the inner accretion disk regions responsible for relativistic FeK$\alpha$ emission.  The broad line in J140700 is consistent with velocities inferred from relativistic FeK$\alpha$ profiles in other AGN \citep[e.g.,][]{nandra07,delacalleperez10}.  However, \textit{apparent} Fe\,K$\alpha$ broadening can also arise from non-relativistic spectral complexity, including continuum curvature from partial-covering absorption, blends of multiple narrow Fe lines, and limited photon statistics \citep[e.g.,][]{reeves+04,nandra07,miller+08,turner+09}.  For this reason, the features cannot be uniquely attributed to relativistic disk emission with the present data.  In Figure~\ref{fig:line_fits} we show the X-ray spectra of three representative sources, illustrating an unresolved line, a possibly moderately broadened line, and a fully broadened line case.


\subsubsection{Fe K$\alpha$ Equivalent Width Completeness Simulations}
\label{results:ew_completeness}

To quantify our sensitivity to Fe K$\alpha$ emission and to interpret both detections and non-detections, we carried out Monte Carlo completeness simulations for each source using unbinned C statistics.  These simulations are used solely to estimate line-recovery completeness as a function of line strength and spectral quality; they do not modify the detection criterion or the statistical uncertainties of the observed spectra.  For every quasar we adopted the best-fit continuum model and generated 100 synthetic spectra with \texttt{fake\_pha}, folded through the appropriate Chandra response files. 

For each source we first produced one hundred continuum-only simulations to establish a null distribution of $\Delta C$ values (where $\Delta C$ is the difference in C-statistic).  Each spectrum in this null set was fit with and without a Gaussian Fe~K$\alpha$ line and the resulting improvement in fit, $\Delta C$, was recorded.  We defined a source-dependent 95\% one-sided detection threshold, $\Delta C_{\rm crit}$, as the value exceeded by only five percent of the null trials.  This procedure yields a fully observation-specific detection threshold that incorporates the photon statistics, background level, instrumental response, and continuum properties of each spectrum.

We then generated one hundred line injected simulations for each of two line classes, with lines injected at the neutral Fe K$\alpha$ emission line energy of 6.4~keV.  For the first set of simulations, we injected an unresolved Gaussian emission line with $\sigma = 0.055$ keV and an equivalent width of $50$ eV.  The adopted width matches the ACIS instrumental broadening of a narrow Fe K$\alpha$ line at 6.4 keV, and the resulting equivalent width lies toward the low end of the distribution observed for narrow cores in local Seyfert galaxies and type~1 AGN, where typical values are $\sim$40–100 eV \citep[e.g.,][]{yaqoob+04,shu+10,ricci+13,ricci+14}.  Injecting relatively weak narrow lines allows us to quantify the minimum Fe K$\alpha$ core strengths that our data are capable of recovering.  A narrow line was counted as recovered when it exceeded $\Delta C_{\rm crit}$ and when the fitted centroid was within $0.15$ keV of 6.40 keV in the rest frame.

For the second set of simulations, we inject broad Gaussian emission lines with $\sigma = 0.212$ keV and an equivalent width of $150$ eV.  These values are representative of relativistic disk lines commonly observed in local Seyfert samples \citep[e.g.][]{guainazzi06,delacalleperez10,patrick+12}, where broad line equivalent widths typically span $100-200$ eV.  This choice provides a conservative but realistic test of our sensitivity to relativistic-disk-broadened Fe K$\alpha$ emission.  A simulated broad line was counted as recovered only when it exceeded $\Delta C_{\rm crit}$, when the fitted centroid was within $0.15$ keV of 6.40 keV in the rest frame, and when the recovered width was consistent with a broad profile, defined as $\sigma > 0.10$ keV. 

The resulting completeness values vary significantly from source to source.  Narrow line completeness ranges from $\sim10-90$\% across the sample, while broad line completeness ranges from only a few percent in the lowest count spectra to more than fifty percent in the best exposed sources.  To evaluate whether our observed Fe K$\alpha$ detection rates are consistent with the sensitivity of the data, we compared the number of detected lines to the number expected based on the completeness simulations.  For each source $i$ we computed the detection probability $p_{i} = f_{\rm int}\,C_{i}$, where $C_{i}$ is the completeness for that source and $f_{\rm int}$ is the assumed intrinsic fraction of AGN that possess a narrow or broad Fe K$\alpha$ line.  The expected number of detections is given by $N_{\exp} = \sum_{i} p_{i}$, and the statistical uncertainty is described by the full Poisson binomial variance, $\sigma_{\rm PB}^{2} = \sum_{i} p_{i}(1-p_{i})$, which properly accounts for sources having different detection probabilities.

Because each completeness value $C_{i}$ is estimated from a finite number of Monte Carlo trials ($N_{\rm line}=100$), there is additional uncertainty associated with the sampling error of the completeness estimates themselves.  For a binomial process with $N_{\rm line}=100$, this uncertainty is $\sigma_{C,i} \approx \sqrt{C_{i}(1-C_{i})/100}$, which propagates to the expected number of detections as $\sigma_{\rm MC}^{2} = \sum_{i} (f_{\rm int}\,\sigma_{C,i})^{2}$.  In practice this term is very small.  For $N_{\rm line}=100$, the Monte Carlo sampling error contributes only a few percent of the total variance and is therefore strongly subdominant to the Poisson binomial term.  The total uncertainty is obtained by summing the two contributions in quadrature.

For the narrow line class we assume that all quasars intrinsically host a narrow Fe K$\alpha$ core ($f_{\rm int}=1$), consistent with large unbiased AGN samples where narrow iron emission is nearly ubiquitous.  Because our narrow line recovery criterion accepts all Fe K$\alpha$ detections with centroids within $0.15$ keV of 6.40 keV, the appropriate observational comparison is the total number of iron line detections in our sample, including unresolved, moderately broadened, and fully broadened cases.  Under these assumptions we predict $N_{\exp} = 8~\pm~2$ Fe K$\alpha$ emission line detections, in good agreement with the six lines actually observed.

For the broad component we assume that half of all quasars possess a relativistic disk line ($f_{\rm int}=0.5$), consistent with the intrinsic $\sim50$\% broad–line fraction inferred from surveys  \citep[e.g.,][]{guainazzi06,nandra07,delacalleperez10,baronchelli18}.  With this assumption, the simulations predict $N_{\exp} = 2~\pm~1$ expected broad line detections, fully consistent with our results (one fully broadened and one possibly moderately broadened).  These comparisons show that the detection statistics predicted by our completeness simulations naturally reproduce the observed incidence of both narrow and broad Fe K$\alpha$ emission.  The agreement indicates that the small number of detected X-ray photons from the quasars in our sample, rather than unusually weak intrinsic iron emission, fully explains the mixture of Fe K$\alpha$ detections and non-detections.

\subsection{Optical-to-X-ray Spectral Index Estimates}

In Table~\ref{table:spectral_fitting} we also report source optical-to-X-ray spectral indices, $\alpha_{ox}$, which are determined between 2500~\text{\AA} and 2~keV, and defined as:

\begin{equation}
	\alpha_{\mathrm{ox}} = \frac{-\log \left( \dfrac{f_{2\,\mathrm{keV}}}{f_{2500\,\text{\AA}}} \right)}{\log \left( \dfrac{\nu_{2\,\mathrm{keV}}}{\nu_{2500\,\text{\AA}}} \right)} .
\end{equation}  

Here, $\nu_{2500\,\text{\AA}}$ is the frequency corresponding to 2500~\text{\AA}, and $\nu_{2\,\mathrm{keV}}$ is the frequency corresponding to 2~keV, with associated flux densities $f_{2500\,\text{\AA}}$ and $f_{2\,\mathrm{keV}}$.  The 2500~\text{\AA} flux densities are derived from SDSS photometry by extrapolating between the extinction-corrected $g$- and $u$-band flux densities using a simple power law\footnote{We use point spread function (PSF) magnitudes for these estimates and apply Milky Way extinction corrections following the reddening maps of \cite{schlafly+11}.}.

The uncertainties in $\alpha_{ox}$ were estimated by propagating the errors in $f_{2500\,\text{\AA}}$ and $f_{2\,\mathrm{keV}}$ (the latter reported in Table~\ref{table:spectral_fitting}).  The $f_{2500\,\text{\AA}}$ errors were derived from uncertainties in the extrapolation of the SDSS flux densities based on the $g$- and $u$-band flux density errors.  To account for the non-simultaneity of the X-ray (Chandra) and optical (SDSS) observations, which are separated by  $\sim$ two decades, we added variability terms to the error budget.  Specifically, we adopt an uncertainty of 0.25 dex in $\log f_{2\,\mathrm{keV}}$ and 0.15 dex in $\log f_{2500\,\text{\AA}}$, consistent with the typical decade-scale variability of quasars with comparable properties to our sample  reported in previous AGN variability studies \citep[e.g.,][]{hook+94,vandenberk+04,macleod+10,middei+17,prokhorenko+24}.

\subsection{Comparison to the Broader Quasar Population}
\label{subsec:gen_pop_comp}

\subsubsection{Photon Index Comparisons}
\label{subsubsec:photon_index_comp}

To investigate whether the X-ray spectral properties of our velocity-offset quasars differ from those of the general quasar population, we compare their X-ray photon indices with those of control samples of normal Type 1 quasars of similar redshift and X-ray luminosity.  For this comparison, we adopt two control samples from which we remove any quasars that also appear in the velocity-offset quasar sample of \citet{eracleous+12} which our Chandra sample is drawn from. 

The first control sample consists of SDSS-selected quasars with serendipitous XMM-Newton coverage \citep{young+09}.  Because this sample shares the same optical selection as our velocity-offset quasars, it provides a well-matched control for assessing potential differences in intrinsic X-ray spectral slopes.  The X-ray photon indices for this control sample were derived from spectral fits performed in the observer-frame $0.5-10~$keV energy range.

The second control sample is drawn from the Chandra Multiwavelength Project (ChaMP; \citealt{green+09}), an X-ray--selected survey of AGN detected in archival Chandra fields.  We include the ChaMP sample to provide an independent comparison with different selection criteria (X-ray versus optical), but observed with the same instrument as our targets.  The X-ray photon indices for this control sample were derived from spectral fits performed in the observer-frame $0.5-8~$keV energy range.  Together, these two complementary control samples allow us to test whether any observed hardening of the X-ray spectra is robust against differences in selection or instrument setup.

To better match control groups to our broad-line–offset quasar sample, we restrict all samples to $0.1 < z < 0.4$ and $43 < \log(L_{2-10\,\mathrm{keV}}) < 45$ (where luminosities are determined in the rest frame).  The latter selection is based upon the luminosity range of our quasar sample, and the redshift selection results in a mean redshift which is consistent across all three samples ($\bar{z}\approx0.27-0.29$).  We also exclude radio-loud AGN\footnote{We compute radio-loudness, R, for our sample as the rest-frame ratio of 5 GHz radio to $B$-band optical flux density (after K-corrections), and adopt R~>~10 as the threshold for radio-loudness \citep{kellerman+94}.  The optical flux densities were derived from the SDSS photometric band that most closely corresponds to the rest-frame $B$-band, and the radio flux densities are extrapolated from X-band observations via in-band spectral indices (Breiding et al., in prep.).} as radio-loud AGN typically exhibit harder X-ray spectra \citep[e.g.,][]{reeves+00}.

\begin{figure}
	\includegraphics[width=\linewidth]{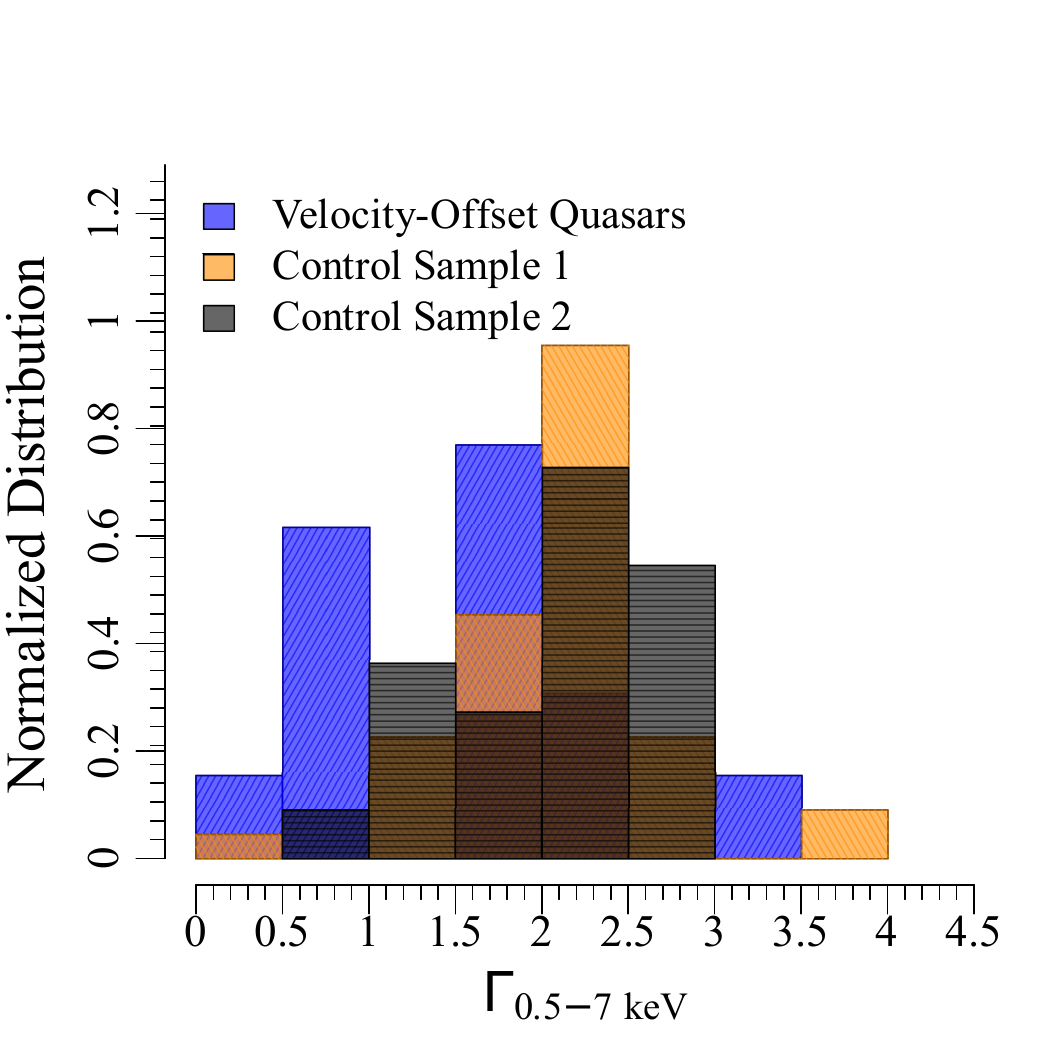}
	\caption{
		Distribution of X-ray photon indices for our sub-sample of quasars with velocity-offset broad emission lines (blue), compared to two control samples shown in orange and black.
		The orange histogram is derived from the SDSS/XMM-Newton Quasar Survey \citep{young+09}, and the black histogram is derived from the ChaMP survey \citep{green+09}.
		All histograms are normalized such that the total area under each distribution is unity, allowing a direct comparison of the shapes of the photon-index distributions independent of sample size.
	}
	\label{fig:index_comps}
\end{figure}

We show histogram comparisons of photon indices between our  velocity-offset quasar sample and the two control samples in Figure~\ref{fig:index_comps}.  For the remainder of this discussion, we refer to the XMM-Newton control sample derived from \cite{young+09} as control sample 1, and the ChaMP control sample derived from \cite{green+09} as control sample 2.  The mean photon index for our velocity-offset quasar sample is 
$\bar{\Gamma}=1.53\pm0.23$\footnote{Quoted uncertainties on $\bar{\Gamma}$ represent the standard error on the mean.  The intrinsic dispersion $\sigma$ characterizes the spread of photon indices within each sample.} 
(with intrinsic dispersion of $\sigma=0.81$), where the control sample means are 
$\bar{\Gamma}_{\rm CS1}=2.12\pm0.09$ and $\bar{\Gamma}_{\rm CS2}=2.04\pm0.14$ 
for control samples 1 and 2 (with corresponding intrinsic dispersions of 
$\sigma_{\rm CS1}=0.59$ and $\sigma_{\rm CS2}=0.65$).  Both control samples have mean photon indices consistent with values found in large surveys of radio-quiet quasars, which typically report $\bar{\Gamma} \simeq 1.8$–$2.0$ \citep[e.g.,][]{reeves+00,george+00,page+05,pioncelli+05}.  In contrast, our velocity-offset quasar sample exhibits a higher fraction of comparatively hard X-ray spectra.  To quantify this difference, we applied Kolmogorov–Smirnov (KS) tests to assess whether the photon-index distributions of our sample and each control sample could be drawn from the same parent population.  We performed one-sided KS tests with the alternative hypothesis that the velocity-offset quasars have systematically smaller (harder) photon indices than the control samples.  The resulting $p$-values are $p = 0.002$ and $p = 0.025$ for control samples~1 and~2, respectively, indicating that the velocity-offset quasars have significantly harder spectra than either control group.  A two-sided KS test between the two control groups yields a p-value of 0.93, indicating the two control groups are consistent with being drawn from the same parent population, namely the broader radio-quiet quasar population with typical broad emission lines.  We also test the robustness of these results by including all radio-loud sources.  The radio-loud AGN fraction for our velocity-offset quasar sample (3/16) was consistent with the fraction found the SDSS-selected control group (6/50), but much less than the radio-loud AGN fraction for control group 2 (31/53).  Based upon binomial statistics with 68\% confidence intervals, the percent of radio-loud AGN in our velocity-offset quasar sample is $19\%^{+15\%}_{-10\%}$.  The corresponding radio-loud AGN percentages for control sample 1 and 2 are $12\%^{+6\%}_{-5\%}$ and $59\%^{+7\%}_{-8\%}$, respectively.  The higher fraction of radio-loud AGN in the X-ray–selected control sample (control~2) is expected, as radio-loud quasars are systematically more X-ray luminous than their radio-quiet counterparts \citep[e.g.,][]{green+95} and are therefore overrepresented in flux-limited X-ray surveys.  Despite the inclusion of radio-loud sources, which generally exhibit systematically harder X-ray spectra, we obtain one-sided KS-test p-values of 0.0003 and 0.04 for control samples~1 and~2, respectively.  These results further support our finding that our broad-line offset quasars possess systematically harder X-ray spectra than the broader quasar population. 

\subsubsection{Optical-to-X-ray Colors}
\label{subsubsec:optical_xray_colors}

\begin{figure}
	\includegraphics[width=\linewidth]{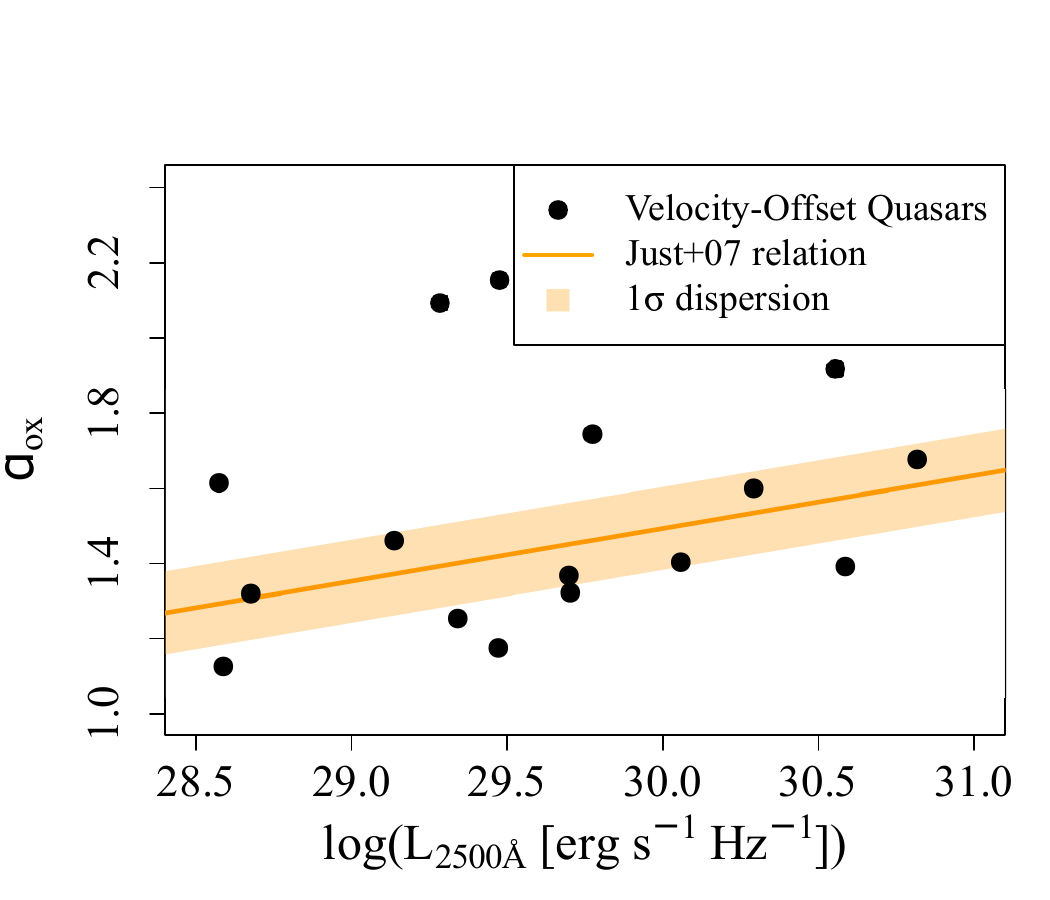}
	\caption{ Optical--to--X-ray spectral slope, $\alpha_{\mathrm{ox}}$, as a function of rest-frame monochromatic UV luminosity, $L_{2500}$, for our velocity-offset quasar sample.  The orange line shows the best-fit $\alpha_{\mathrm{ox}}$--$L_{2500}$ relation from \citet{just+07}, while the shaded region denotes the $\pm1\sigma$ dispersion determined solely by the intrinsic scatter in the \citet{just+07} sample fit.  Black points represent our sample, which shows no significant systematic deviation from the canonical relation, consistent with typical Type~1 quasar optical--to--X-ray colors.
	}
	\label{fig:aox_comps}
\end{figure}

As a further test for possible differences in X-ray emission relative to normal quasars, we compared the optical–to–X-ray spectral slopes of our sample with the $\alpha_{\mathrm{ox}}$–$L_{2500}$ relation of \citet{just+07}.  The \citet{just+07} relation is given by: 
\[
\alpha_{\mathrm{ox}} = (-0.140 \pm 0.007)\,\log L_{2500} + (2.705 \pm 0.212),
\]
and has an intrinsic scatter of $\sigma_{\mathrm{int}} \simeq 0.11$. 
For each quasar we computed the residual $\Delta\alpha_{\mathrm{ox}} = \alpha_{\mathrm{ox,obs}} - \alpha_{\mathrm{ox,exp}}$ to quantify deviations from the expected relation. 
Across the 16 quasars in our sample, we find a mean residual of $\langle\Delta\alpha_{\mathrm{ox}}\rangle = 0.10$, indicating that, on average, our sources are marginally X-ray strong relative to the canonical quasar population.  However, this offset is not statistically significant: a one-sample $t$-test yields $\mathrm{p = 0.21}$, consistent with the null hypothesis that the residuals are centered on zero.  Additionally, excluding the two obvious outliers which correspond to the extremely low count sources with large uncertainties (J121113 and J130524), significantly reduces the X-ray enhancement to much less than the intrinsic scatter: $\langle\Delta\alpha_{\mathrm{ox}}\rangle = 0.01$.  Approximately half of the quasars (excluding the two clear outliers) fall within the intrinsic $1\sigma$ scatter of the \citet{just+07} relation (shown as the orange band in Figure~\ref{fig:aox_comps}), suggesting that the dispersion in our sample may be somewhat larger than in large, homogeneous quasar samples. 
Such additional variance is expected given the small sample size, non-simultaneity of the optical and X-ray observations, and potential source heterogeneity. 

\subsection{Black Hole Mass Estimates}
\label{results:mass_estimates}

\begin{table*}
	\caption{Inferred Black Hole Mass and Eddington Ratios} 
	\label{table:phys_prop}
	\begin{threeparttable}
		\begin{tabular}{lrrrrrrr} 
			\hline
			Source&log~$(\lambda^{\alpha_{ox}})$&log~$(\lambda^{\Gamma})$&log~$\left(\frac{M^{M_{\bullet}-\sigma_{*}}_{\bullet}}{M_{\odot}}\right)$&log~$\left(\frac{M^{\alpha_{ox}}_{\bullet}}{M_{\odot}}\right)$&log~$\left(\frac{M^{\Gamma}_{\bullet}}{M_{\odot}}\right)$&log~$\left(\frac{M^{f.p.}_{\bullet}}{M_{\odot}}\right)$&log~$\left(\frac{M^{\sigma^{2}_{NXS}}_{\bullet}}{M_{\odot}}\right)$\\
			Name&&&&&&&\\
			\hline
			J015530&$-1.1\pm0.5$&$-0.8\pm0.9$&$8.1\pm1.0$&$8.4\pm0.5$&$8.2\pm1.0$&$5.2\pm1.4$&$>6.3$\\
			J020011&$-1.0\pm0.5$&$-0.8\pm0.9$&$9.2\pm0.9$&$8.6\pm0.5$&$8.5\pm1.0$&...&$>6.2$\\
			J092712&$-0.3\pm0.5$&$-1.6\pm0.8$&$7.6\pm1.0$&$7.5\pm0.6$&$8.4\pm0.9$&...&...\\
			J093844&$-1.1\pm0.5$&$-1.9\pm0.8$&$8.8\pm1.0$&$8.4\pm0.6$&$8.9\pm0.9$&$6.3\pm1.4$&$>6.2$\\
			J095036&$-0.9\pm0.5$&$-0.7\pm0.9$&$8.8\pm0.9$&$7.4\pm0.5$&$7.3\pm1.0$&$8.7\pm1.3$&...\\
			J102839&$-0.6\pm0.5$&$-0.5\pm1.0$&$8.1\pm0.9$&$8.2\pm0.6$&$8.1\pm1.0$&...&...\\
			J110556&$-1.2\pm0.5$&$-1.6\pm0.8$&$9.5\pm0.9$&$8.8\pm0.6$&$9.1\pm0.9$&...&...\\
			J113651&$-1.4\pm0.5$&$-0.7\pm0.9$&$8.2\pm0.9$&$8.1\pm0.5$&$7.6\pm1.0$&$5.0\pm1.3$&$>6.1$\\
			J121113&$>-1.1$&$0.4\pm3$&$9.0\pm0.9$&$<9.1$&$7.5\pm3$&$6.3\pm1.8$&...\\
			J124551&$-0.5\pm0.5$&$-2.3\pm0.7$&$8.4\pm0.9$&$7.9\pm0.5$&$9.0\pm0.8$&...&$>6.6$\\
			J125142&$-1.3\pm0.5$&$-1.0\pm0.9$&$8.6\pm0.9$&$8.9\pm0.5$&$8.7\pm0.9$&$5.3\pm1.4$&$>6.3$\\
			J130534&$>-0.9$&$-0.3\pm1$&$8.5\pm0.9$&$<8.6$&$6.2\pm4$&$5.5\pm2$&...\\
			J140251&$-0.6\pm0.5$&$-0.7\pm0.9$&$8.6\pm0.9$&$8.5\pm0.5$&$8.2\pm1.0$&...&...\\
			J140700&$-1.1\pm0.5$&$-1.3\pm0.8$&$8.7\pm1.0$&$6.8\pm0.5$&$7.2\pm0.9$&$9.8\pm1.3$&...\\
			J151443&$-0.7\pm0.5$&$-1.2\pm0.9$&$8.5\pm0.9$&$9.3\pm0.5$&$9.4\pm0.9$&$9.2\pm1.6$&$>6.5$\\
			J165118&$-0.7\pm0.5$&$-1.5\pm0.8$&$9.2\pm0.9$&$8.6\pm0.5$&$9.0\pm0.9$&...&...\\
			\hline
		\end{tabular}
		\begin{tablenotes}	
			\item[$\dagger$] All upper and lower limits are 95\% limits.  We do not report black hole masses using the black hole fundamental plane relation for sources not detected in the radio.  Similarly, we do not report black hole masses constrained from normalized excess variance for sources in which we do not complete the variability analysis.
		\end{tablenotes}
	\end{threeparttable}
\end{table*}

In this section we derive black hole masses using several different techniques.  We begin with estimates based on Eddington ratios inferred from X-ray scaling relations.  We then present masses obtained from X-ray variability measurements, followed by those derived from the fundamental plane of black hole activity.  We also estimate black hole masses using the $M_{\bullet}$--$\sigma_\ast$ relation, where the velocity dispersion is approximated from the widths of narrow emission lines.  Finally, we compare the results from all methods using both object-level examples and statistical comparisons across the full sample to assess their overall consistency.  All scaling relations used in this section are listed in Appendix~\ref{app:scaling_relations}, along with their coefficient uncertainties and intrinsic scatter.  In all of our mass estimates we propagate every available source of uncertainty, including measurement errors, uncertainties in the scaling relation coefficients, and the intrinsic scatter of each relation.  The resulting mass measurements are all listed in Table~\ref{table:phys_prop}, along with the measured Eddington ratios.

The different mass estimators employed here rely on overlapping observables and are therefore best interpreted as complementary diagnostics.  
In particular, the accretion-based estimates derived from $\Gamma$ and $\alpha_{ox}$, as well as the fundamental-plane relation, all depend on X-ray spectral properties and are thus not statistically independent.  
Moreover, each method incorporates intrinsic scatter from empirical scaling relations that does not represent a measurement error in the usual statistical sense.  
For these reasons, the black hole mass estimates derived from different techniques should be interpreted as complementary probes of black hole mass rather than as independent measurements of a single underlying quantity.  The implications of these differences under the binary SMBH interpretation are discussed in detail in Section~\ref{res:mass_comparison}.

A primary motivation for employing multiple mass estimators is to test a specific prediction of the binary SMBH interpretation.  In hydrodynamic simulations of accreting binaries, the lower–mass secondary typically dominates the accretion luminosity, implying that accretion-based methods trace the mass of the active component, whereas host–galaxy–based methods such as the $M_{\bullet}$--$\sigma_{\ast}$ relation may better approximate the total black hole mass of the system.  If this picture applies, the $M_{\bullet}$--$\sigma_{\ast}$ masses should on average exceed the accretion-based estimates.  We evaluate this expectation in Section~\ref{res:mass_comparison}.

\subsubsection{Eddington Ratio Estimates}
\label{res:eddington}

We estimated the Eddington ratios of our quasars using two different X-ray-based indicators.  First, we applied the empirical correlation between the hard X-ray photon index and Eddington ratio from \citet[][their Equation~3]{shemmer+08}.  In that work the photon indices were obtained by fitting rest–frame energies above 2~keV\footnote{Our photon indices are measured over the observer-frame 0.5--7~keV band.  For our sources which range in redshift from $z\!\approx\!0.1$--$0.7$, this corresponds to rest-frame energies of roughly 0.6--12~keV, which samples the same hard X-ray continuum that defines the rest-frame 2--10~keV band used by \citet{shemmer+08}.  This observer-frame energy range is standard for Chandra spectral fitting of unobscured type~1 AGN.}.  Second, we used the $\alpha_{\mathrm{ox}}$--Eddington ratio relation of \citet[][their Equation~16]{lusso+10}, adopting their formulation that treats $\alpha_{\mathrm{ox}}$ as the independent variable.  

For each object we evaluated both scaling relations using the best-fit photon indices from our X-ray spectral modeling and the measured $\alpha_{\mathrm{ox}}$ values.  The Eddington ratios inferred from $\Gamma$ and from $\alpha_{\mathrm{ox}}$ are broadly consistent within the expected uncertainties of the relations.  The accretion rates for our quasar sample are typically of order ten percent of the Eddington limit, with most sources spanning $\sim5-50$\%.  

\subsubsection{Black Hole Mass Determination from Eddington Ratios}
\label{res:mass_estimates_edd}

We translated our Eddington ratio estimates into black hole masses by combining them with the intrinsic rest-frame $2$--$10$~keV luminosities, $\mathrm{L_{2\text{--}10\,keV}}$, derived from our spectral fitting and the bolometric correction–Eddington ratio relation from \citet{lusso+12}.  We used the regression in their Table~3 that treats the Eddington ratio as the independent variable when predicting the hard X-ray bolometric correction.  This relation provides the bolometric correction factor as a function of $\lambda$, allowing the bolometric luminosity to be estimated directly from our X-ray data as $\mathrm{L_{bol} = K\,L_{2\text{--}10\,keV}}$, where $K$ is the corresponding bolometric correction factor.

Given the bolometric luminosity and Eddington ratio, the black hole mass follows from the relation:
\begin{equation}
	M_{\mathrm{\bullet}} = 
	\frac{L_{\mathrm{bol}}}{\lambda \, (1.26\times10^{38}\ \mathrm{erg\ s^{-1}\ M_{\odot}^{-1}})} .
\end{equation}
We computed masses separately for the Eddington ratios inferred from the X-ray photon indices and the optical-to-X-ray spectral indices.  The two mass estimates agree within the expected systematic uncertainties that arise from the intrinsic scatter in the underlying correlations.  The resulting black hole masses mostly fall in the range ${\sim}10^{7.5}$--$10^{9}\,M_{\odot}$, consistent with expectations for luminous quasars at these redshifts.

\subsubsection{X-ray Variability Constraints on Black Hole Mass}
\label{res:mass_var}

\begin{table*}
	\centering
	\caption{Parameters \& Results of Light Curve Analysis} 
	\label{table:excess_var}
	\begin{threeparttable}
		\begin{tabular}{lrrrrrr} 
			\hline
			Source&Time Bin&Obs.  Length&Calibration&$\mathrm{N_{bins}}$&$\mathrm{\sigma^{2}_{NXS}}$&Correction\\
			Name&Size&&&&&Factor\\
			&(s)&(ks)&(ks)&&&\\
			\hline
			J015530&465&17.9&20&39&$<0.010$&1.15\\
			J020011&765&21.3&20&29&$<0.016$&0.98\\
			J093844&891&19.1&20&22&$<0.015$&1.13\\
			J113651&698&18.2&20&27&$<0.018$&1.18\\
			J124551&1799&74.0&40&42&$<0.014$&0.56\\
			J125142&250&19.2&20&39&$<0.014$&1.04\\
			J151443&250&30.4&40&123&$<0.007$&1.32\\
			
			\hline
		\end{tabular}
		\begin{tablenotes}	
			\item[] Upper limits on $\mathrm{\sigma^{2}_{NXS}}$ are 95\% limits.
			\item[] All light curve properties are determined in the rest frame for our sources.
		\end{tablenotes}
	\end{threeparttable}
\end{table*}

Using constraints on short-timescale variability derived from the ACIS light curves constructed from the observations in Tables~\ref{table:cxo_obs} and \ref{table:cxo_arch}, we estimated black hole masses via the empirical correlation between normalized excess variance and black hole mass established by \citet{ponti+12}.  
Table~\ref{table:excess_var} summarizes the light curve properties for all quasars with sufficient temporal sampling, including the adopted time-bin sizes, observation lengths, frequency-band correction factors, and resulting upper limits on the normalized excess variance.

For each quasar we measured the rest-frame $2$--$10$~keV normalized excess variance, $\sigma^{2}_{\mathrm{NXS}}$, following the methodology described in Section~\ref{subsec:light_curve_analysis}.  For all seven quasars with adequately sampled light curves, we find no significant intrinsic variability.  In every case the $1\sigma$ confidence interval of $\sigma^{2}_{\mathrm{NXS}}$ includes zero, and we therefore report $95\%$ upper limits on $\sigma^{2}_{\mathrm{NXS}}$ (Table~\ref{table:excess_var}).  

When the upper limits on $\sigma^{2}_{\mathrm{NXS}}$ are propagated through the $\sigma^{2}_{\mathrm{NXS}}$–$M_{\mathrm{\bullet}}$ scaling relation, they translate into lower limits on the black hole masses.  The resulting mass constraints imply $\log(M_{\mathrm{\bullet}}/M_{\odot}) \gtrsim 6.2$--$6.6$, consistent with the expectation that the absence of detectable short-timescale variability indicates relatively large black hole masses.  These limits are generally well below the black hole masses inferred using the other techniques presented in this paper.  Under the binary SMBH hypothesis, these limits would constrain the mass of the actively accreting component; the broader implications are considered in Section~\ref{res:mass_comparison}.

\subsubsection{Black Hole Masses from the Fundamental Plane of Black Hole Activity}
\label{res:mass_fp}

We also estimated black hole masses using the fundamental plane of black hole activity, an empirical relation linking compact radio emission, hard X-ray luminosity, and black hole mass across the full black-hole mass spectrum.  We adopt the calibration of \citet{gultekin+19}, which is specifically derived using core 5~GHz radio luminosities and does not include extended emission components.  This choice ensures that the relation reflects emission associated with the central engine rather than kiloparsec-scale jets or lobes.  The full scaling relation and its coefficients are listed in Appendix~\ref{app:scaling_relations}.

We use the high–angular–resolution VLA X-band A-configuration data described in \citet{breiding+21}, with the full dataset to be presented in a future publication (Breiding et al., in prep.), to isolate compact core emission and derive rest-frame 5~GHz luminosities consistent with the requirements of the \citet{gultekin+19} fundamental-plane calibration.  These data isolate compact emission on sub-kiloparsec scales and effectively suppress contamination from extended radio structures.  We derived rest-frame 5~GHz luminosities by using measured in-band spectral indices to apply K corrections to the observed 10 GHz X-band data.  The radio luminosities were combined with the intrinsic rest-frame $2$--$10$~keV luminosities obtained from our spectral fitting (Table~\ref{table:spectral_fitting}) to infer black hole masses.  Nine quasars in our sample have the requisite compact radio detections and therefore permit fundamental-plane mass estimates, as summarized in Table~\ref{table:phys_prop}.  Although we account for the measurement uncertainties in the radio and X-ray luminosities, the total error budget for the fundamental-plane masses is dominated by the large intrinsic scatter of $\sim$~1 dex in the relation itself.  Where available, the fundamental-plane masses tend to lie toward the lower end of the range spanned by our other estimates.  For sources without compact radio detections we do not report fundamental-plane mass estimates.

\subsubsection{Black Hole Masses from the \texorpdfstring{$M_{\bullet}$--$\sigma_\ast$}{M-sigma} Relation}
\label{res:mass_msigma}

Our final black hole mass measurement technique relies on the $M_{\bullet}$--$\sigma_\ast$ relation, where we use the widths of narrow emission lines as proxies for the stellar velocity dispersion of the host galaxies.  For each quasar we adopted the multi-epoch [O\,\textsc{iii}]\,$\lambda5007$ line-width measurements reported by \citet{runnoe+17}, taking the mean of their measurements as the representative FWHM and the epoch-to-epoch standard deviation as the corresponding uncertainty on the line width.  These averaged FWHM values were then converted to stellar velocity dispersions using the empirical calibration of \citet{brotherton+15}, which provides a direct mapping between FWHM([O\,III]) and $\sigma_\ast$ without assuming a Gaussian profile for the narrow-line emission.  Finally, we applied the \citet{tremaine+02} $M_{\bullet}$--$\sigma_\ast$ relation (see Appendix~\ref{app:scaling_relations}) to obtain black hole mass estimates for each quasar.  We emphasize that the uncertainties on the $M_{\bullet}$--$\sigma_\ast$ masses should be interpreted with caution.  
The scatter among the narrow-line widths reflects physical differences in the kinematics of the narrow-line region, which can be influenced by outflows, radiation pressure, and ionization stratification, rather than purely statistical measurement errors.  
As a result, the epoch-to-epoch dispersion in the narrow-line widths does not strictly represent a measurement uncertainty on $\sigma_\ast$ in the usual sense.  
In practice, the total uncertainty on the $M_{\bullet}$--$\sigma_\ast$ masses is therefore dominated by systematic effects and by the intrinsic scatter of the underlying scaling relations, rather than by the precision of the narrow-line measurements themselves.

Because the $M_{\bullet}$--$\sigma_\ast$ method reflects the gravitational potential of the host galaxy, it may approximate the total black hole mass in a bound binary system, although this interpretation relies on the assumption that the narrow-line region responds to the combined potential of both components.  In contrast, the accretion-based methods discussed earlier primarily reflect the mass of the SMBH that dominates the accretion luminosity, which simulations commonly show to be the lower-mass secondary in binary systems.  This distinction becomes important in Section~\ref{res:mass_comparison}, where we compare all mass estimators in the context of the binary SMBH hypothesis.

\subsubsection{SDSS / Virial Black Hole Mass Estimates}
\label{sec:sdss_virial}

For each quasar we also include the single–epoch virial black hole mass estimates reported in the SDSS DR7 quasar catalog of \citet{shen+11} when comparing the various black hole mass estimates used in this work in Section~\ref{res:mass_comparison}.  This provides an additional estimate for the mass of the accreting black hole responsible for the optical broad-line emission, based upon the optical broad-line emission properties.

We adopt the fiducial SDSS virial black hole masses from \citet{shen+11}, who assign one consistent virial estimate per quasar based on the most reliable broad emission line available at that redshift.  The black hole mass uncertainties reported by \citet{shen+11} include only the measurement errors from the line-width and continuum fits.  For our analysis, we combine these reported measurement errors in quadrature with the uncertainty associated with the virial scaling relation.  Following \citet{shen+13b}, who argue that single–epoch virial masses are accurate to only $\sim$0.5~dex, we adopt a total uncertainty of 0.5~dex when applying the single–epoch scaling relation.

We note, however, that single–epoch virial mass estimates should be treated with particular caution for velocity-offset quasars.  
These estimators are calibrated for typical, isolated quasars and assume a virialized BLR.  
If the actively accreting black hole resides in a binary, the BLR may be truncated or dynamically perturbed, which can broaden the emission lines and bias virial masses high \citep[e.g.,][]{runnoe+17}.  
In this case, systematic offsets between virial and accretion-based masses may reflect differences in BLR structure rather than inconsistencies among the mass estimation techniques.

\subsection{Comparison of Black Hole Mass Estimates}
\label{res:mass_comparison}

A key motivation for obtaining multiple black hole mass estimates is to test whether the host–galaxy–based $M_{\bullet}$–$\sigma_{\ast}$ masses are systematically higher than the masses inferred from techniques that depend on the actively accreting component.  In the binary SMBH scenario, hydrodynamic simulations show that the lower–mass secondary typically dominates the accretion rate and therefore the observed X-ray and optical/UV emission.  If this picture applies to our quasars, then the accretion-based mass estimates would trace the presumably lower-mass active component, whereas the $M_{\bullet}$–$\sigma_{\ast}$ relation would approximate the total black hole mass of the system.  Under this hypothesis, $M_{\bullet}$–$\sigma_{\ast}$ masses should on average exceed the accretion-based estimates.  

We began by testing whether the different mass--estimation techniques yield systematically different masses across the sample. We used a linear mixed--effects model \citep[e.g.,][]{gelman_hill_06} because each quasar has multiple method-dependent mass estimates that are not independent but are correlated by virtue of referring to the same underlying black hole. In the mixed--effects model, the mass--estimation method enters as a fixed effect, since we want to estimate and compare the average mass given by each technique. The quasars themselves are modeled as random effects, allowing each object to have its own baseline mass while still testing for systematic differences between methods. The model was fitted using inverse--variance weights so that mass estimates with smaller quoted uncertainties contribute more strongly to the fit, while recognizing that these uncertainties incorporate both measurement error and intrinsic scatter in the underlying scaling relations rather than purely statistical noise. We then tested the statistical significance of the fixed effect associated with the mass--estimation method using an $F$-test in the mixed--effects framework\footnote{All statistical modeling was performed in \texttt{R} \citep{Rcore}, using the \texttt{lme4} package \citep{lme4} to fit the linear mixed--effects model and the \texttt{lmerTest} package \citep{lmerTest} to compute the $F$ statistic and associated $p$-value for the fixed effect.}. In our case, this test yields $F=8.84$ with an associated $p$-value of $1.6\times10^{-5}$, indicating that the choice of mass--estimation method has a statistically significant effect on the inferred masses.

Inspection of the estimated marginal means from the mixed--effects model clarifies the pattern underlying the significant method effect.  The five methods yield the following marginal means with standard errors:
$8.19 \pm 0.14$ for the $\alpha_{\mathrm{ox}}$–based method,
$7.03 \pm 0.39$ for the fundamental plane,
$8.38 \pm 0.22$ for the photon index–based method,
$8.81 \pm 0.11$ for the SDSS virial method,
and $8.63 \pm 0.20$ for the $M_{\bullet}$--$\sigma_{\ast}$ method.
Together, these values indicate that the fundamental–plane method tends to yield systematically lower masses than the other techniques, while the SDSS virial method yields the highest characteristic masses in the sample.

To determine which differences are statistically significant, we compared the model-estimated mean masses for each technique using standard $t$-tests applied to the mixed-model predicted means, with Holm-adjusted $p$-values to account for multiple comparisons.
The fundamental–plane method produced significantly lower masses than the $\Gamma$–based method ($p=0.039$), the SDSS virial method ($p=0.0023$), and the $M_{\bullet}$--$\sigma_{\ast}$ method ($p=0.0106$)\footnote{The remaining comparison of the fundamental plane to the $\alpha_{\mathrm{ox}}$–based method returned $p=0.057$, which we do not count as statistically significant}.
In addition, the SDSS virial masses were significantly higher than the $\alpha_{\mathrm{ox}}$–based masses ($p=0.0002$).  

\begin{figure*}
	\centering
	\includegraphics[width=\textwidth]{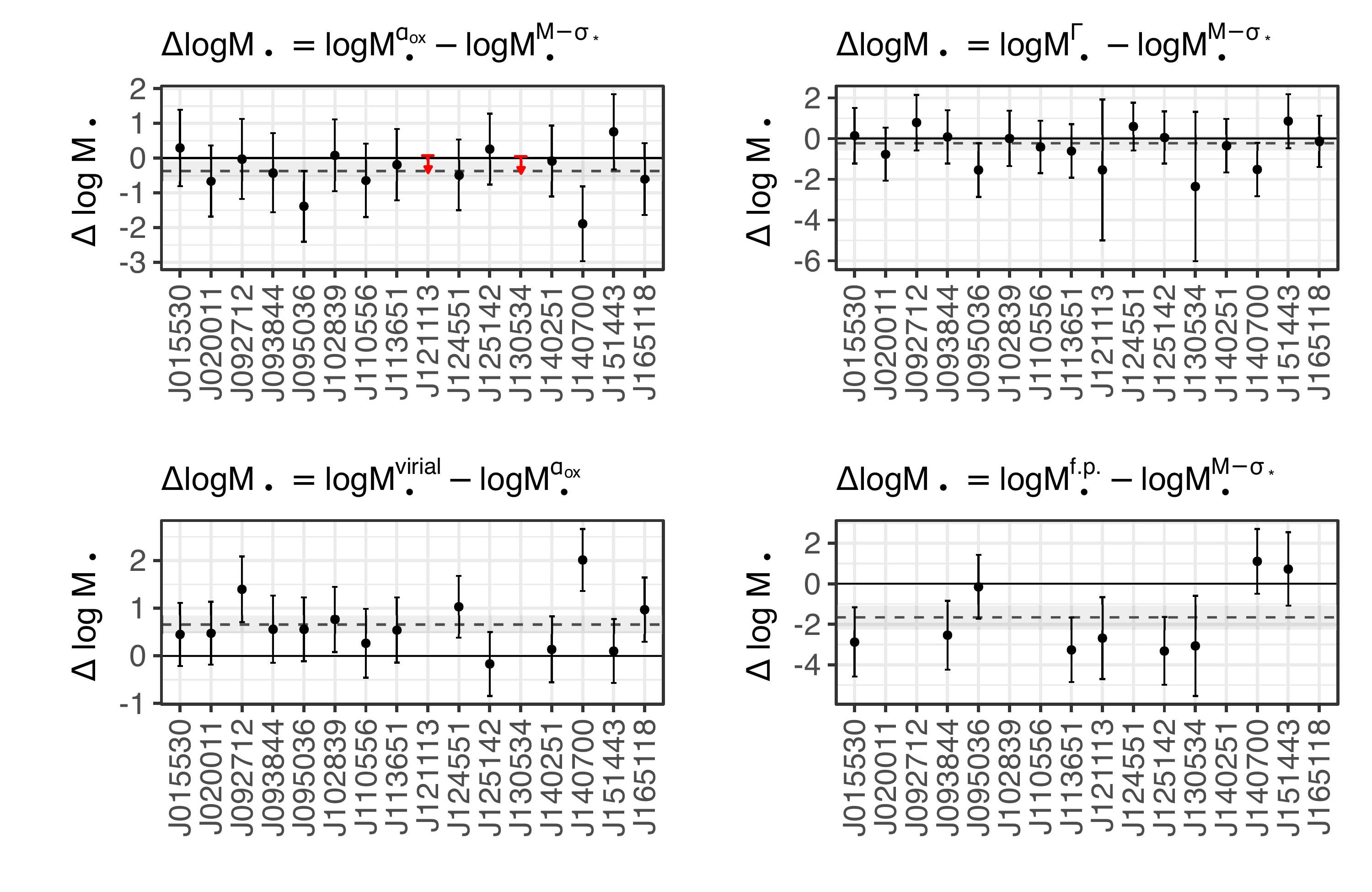}
	\caption{
		Pairwise comparisons of black hole mass estimates derived from different methods for the 16 quasars in our sample.
		Each panel shows $\Delta \log M_{\bullet}$ for individual sources, ordered by increasing right ascension, with $1\sigma$ uncertainties propagated in quadrature from the two mass estimates being differenced.
		From top left to bottom right, the panels compare
		$\log M_{\bullet}^{\alpha_{\mathrm{ox}}}$ to $\log M_{\bullet}^{M_{\bullet}\!-\!\sigma_{\ast}}$,
		$\log M_{\bullet}^{\Gamma}$ to $\log M_{\bullet}^{M_{\bullet}\!-\!\sigma_{\ast}}$,
		$\log M_{\bullet}^{\mathrm{virial}}$ to $\log M_{\bullet}^{\alpha_{\mathrm{ox}}}$,
		and $\log M_{\bullet}^{\mathrm{f.p.}}$ to $\log M_{\bullet}^{M_{\bullet}\!-\!\sigma_{\ast}}$..
		The solid horizontal line marks $\Delta \log M_{\bullet} = 0$.
		The dashed gray line indicates the inverse-variance–weighted mean difference in each panel, with the shaded band showing the corresponding $\pm1\sigma$ uncertainty (standard error of the mean).
		These mean offsets are included as a visual guide to systematic trends across the sample and do not represent combined or preferred mass estimates.
		Red arrows denote sources for which the $\alpha_{\mathrm{ox}}$–based mass is an upper limit, resulting in an upper limit on $\Delta \log M_{\bullet}$ in the relevant panels.
	}
	\label{fig:mass_comp_panel}
\end{figure*}

While the mixed–effects analysis establishes that the inferred black hole masses depend significantly on the estimation method, it does not convey how these offsets manifest on an object-by-object basis.  To illustrate the structure of these differences across the sample, Figure~\ref{fig:mass_comp_panel} presents a set of targeted, pairwise comparisons between selected mass estimators.  Rather than showing absolute masses, each panel plots the difference between two estimators for each source, allowing systematic offsets to be identified while minimizing object-to-object variations in the overall mass scale.  The specific pairings shown are chosen to reflect the most salient trends revealed by the statistical analysis, including the behavior of accretion-based estimators relative to $M_{\bullet}$--$\sigma_{\ast}$ masses, the tendency for SDSS virial masses to exceed X-ray–based estimates, and the systematically lower masses inferred from the fundamental plane.  In each panel, the horizontal line marks the inverse-variance–weighted mean difference and serves as a descriptive guide to the typical offset between the two methods across the sample.  Together, these comparisons provide a compact visual summary of the method-dependent mass offsets discussed above and place the statistical results in the context of individual sources.

\section{Discussion}
\label{discussion}

As outlined in Section~\ref{intro}, several hypotheses have been proposed to explain the velocity–offset broad optical emission lines observed in our quasars, including dynamical motion of a binary SMBH, a recoiling SMBH, and BLR gas kinematics within a single-SMBH system.  A major goal of this study is to test whether the X-ray properties of our velocity-offset quasar sample support or challenge any of these competing scenarios.  While our results do not exclude any of the proposed explanations outright, they do help to narrow the range of plausible physical interpretations.

\subsection{Hard X-ray Continua in Velocity–Offset Quasars}

Relative to well-matched control samples, our quasars exhibit significantly harder X-ray continua (\S~\ref{subsubsec:photon_index_comp}), with a mean photon index $\Gamma = 1.53\pm0.23$,  below the typical 
$\Gamma \approx 1.8$–2.0 observed in radio-quiet Type~1 quasars.  Binary accretion flows can in principle alter the hard X-ray spectrum, most notably through shock heating where gas streams from the circumbinary disk strike the minidisks.  These shocks generate an additional hard X-ray component whose characteristic temperature is set by the orbital velocity at the stream–impact radius and thus depends sensitively on the binary separation and mass ratio.  A useful benchmark comes from the stream–minidisk shock simulations of \citet[][]{roedig+14}.  
For a binary with separation $a = 100\,R_{\rm g}$, these models yield post-shock temperatures corresponding to a Wien-like excess peaking at $E_{\rm peak} \sim 100$~keV.  Following the approach of \citet{foord+17}, we use the stream–shock scalings of \citet{roedig+14} to relate the post–shock temperature to the binary separation, a.  
For an equal-mass binary (which is most consistent with the mass constraints presented in \S~\ref{res:mass_comparison}), since $T_{\rm ps} \propto a^{-1}$, the corresponding Wien peak scales as:
\begin{equation}
	E_{\rm peak} \approx 100~{\rm keV}
	\left(\frac{a}{100\,R_{\rm g}}\right)^{-1}.
\end{equation}

To move the shock peak into the $E_{\rm peak} \sim 10$~keV regime where its low-energy tail could meaningfully harden our observer-frame 0.5--7~keV spectra, we require:
\begin{equation}
	a \approx 10^3\,R_{\rm g}.
\end{equation}
Notably, separations of this order are also those independently expected for the
velocity-offset quasars in the \citet{eracleous+12} sample under the binary-SMBH
interpretation, based on their observed velocity offsets and long-term stability \citep{pflueger+18}.  For a $10^{8}\,M_{\odot}$ black hole, $R_{\rm g}=GM/c^2 \simeq 4.8\times10^{-6}$~pc, so a separation of $a \sim 10^3\,R_{\rm g}$ corresponds to $\sim 4.8\times10^{-3}$~pc ($\approx 6$ light-days).  
At such separations, a $\sim$10~keV shock component with luminosity comparable to the coronal continuum could plausibly harden the effective photon index from $\Gamma\simeq1.8$ to $\Gamma\simeq1.5$.  
However, several 
more conventional explanations can also account for the observed 
spectral hardening.

Another possibility for the harder X-ray continua we observe is modest, 
unmodeled neutral absorption.  Neutral hydrogen has strong photoelectric 
opacity at soft energies, so even low intrinsic columns can suppress the 
0.5--2~keV flux and introduce subtle curvature
that is difficult to distinguish from an intrinsically harder,
unabsorbed continuum.  Although 
our fits required intrinsic neutral absorption for only a few quasars (see
Section~\ref{subsec:spectral_analysis} for our methodology and
Table~\ref{table:spectral_fitting} for the specific cases), 
columns below our detection threshold could still bias the inferred photon 
index.

Warm absorbers and partial-covering structures provide a distinct 
mechanism for apparent spectral hardening.  In ionized gas, the relevant 
soft X-ray opacity is dominated not by hydrogen but by metals such as 
O, Ne, Mg, Si, S, and Fe, whose K- and L-shell edges and associated 
line complexes produce curvature across $\sim0.5-2$~keV that blends into 
a smooth continuum at Chandra’s CCD resolution 
\citep[e.g.,][]{reynolds97,krongold+03,turner+09}.  A further possibility is X-ray reflection from the
accretion disk or from distant Compton-thick material, which produces a
neutral Fe\,K$\alpha$ line and a Compton reflection hump peaking at
$\sim$20--30~keV; when such a component is present, the added hard flux
above a few keV can flatten the photon index inferred from a single
power-law fit \citep{george+91,ricci+11}.  Deeper X-ray observations
would better constrain any soft-band curvature from neutral or warm
absorbers, while higher spectral resolution would allow us to detect
individual ionized absorption features.  Conversely, observations
extending above 10~keV (e.g., with NuSTAR) can directly test the
reflection scenario by searching for a Compton hump and enhanced
Fe\,K$\alpha$ equivalent widths, features that are unconstrained by our
current data. 

In this context, our results are consistent with recent theoretical work demonstrating that binary SMBH accretion can be difficult to diagnose using CCD-resolution X-ray spectra alone. In particular, \citet{malewicz+25} showed that relativistic reflection spectra from binary systems with multiple mini-disks can be acceptably fit with standard single-AGN models in the low-count regime, even when the underlying spectra arise from composite accretion flows with disparate ionization states. As a result, hard photon indices or apparently anomalous reflection parameters do not uniquely signal either the presence or absence of a binary SMBH. Instead, such features reflect the fundamental degeneracy between intrinsic continuum shape, absorption, and reflection when only moderate signal-to-noise X-ray spectra are available. Our Chandra observations probe this same observational regime, reinforcing the conclusion that subtle spectral hardening alone cannot serve as a definitive discriminator between binary and single-SMBH scenarios.

Intrinsic coronal variability provides another plausible origin for the 
hard photon indices.  In thermal Comptonization models, the X–ray 
spectral slope depends primarily on the electron temperature and 
optical depth of the corona, with hotter or more optically thick plasmas 
producing intrinsically harder continua \citep[e.g.,][]{sunyaev+80,lightman+87,zdziarski+96}.  Coronal compactness provides an additional 
constraint: increasing compactness enhances pair production, which in 
turn regulates the electron temperature and modifies the resulting 
spectral shape \citep[e.g.,][]{stern+95,fabian+15}.  In this scenario, the 
observed hardness reflects differences in coronal heating or geometry 
rather than absorption or reflection, and such differences are consistent 
with both single–SMBH coronae and the distinct coronal environments 
expected in binary accretion flows.

Although our quasars exhibit somewhat hard X-ray spectra, their 
optical–to–X-ray spectral energy distributions (SEDs) remain consistent with the broader quasar 
population (see Section~\ref{subsubsec:optical_xray_colors}).  Thus, any binary SMBH interpretation must allow for modest hardening of 
the X-ray continuum while otherwise preserving a normal quasar UV–to–X-ray SED. 
Alternatively, if future observations reveal modest intrinsic absorption, 
the location and kinematics of that gas could link the X-ray hardening to 
scenarios in which the BLR is embedded in, or accompanied by, a 
structured inflow or outflow that may also produce the velocity-offset optical 
broad lines.  Future high–resolution X-ray spectroscopy could also test for ultrafast
outflows (UFOs) through Fe K-shell absorption features from $7-10$~keV 
\citep[e.g.,][]{tombesi+10,tombesi+13}.  Such outflows are assumed to be associated with accretion disk winds which originate on scales 
comparable to (or interior to) the BLR, and in some systems have been shown 
to drive large–scale molecular outflows \citep{tombesi+15}.  Detecting or constraining UFOs in our quasars would therefore help to 
assess whether powerful disk winds are present, providing additional 
context for outflow-based interpretations of the velocity-offset broad 
lines.  Conversely, detection of strong reflection features, such as a pronounced 
Compton hump or large Fe\,K$\alpha$ equivalent widths, would instead 
indicate that part of the hardening arises from reprocessing rather than 
from the intrinsic continuum shape, thereby placing geometric constraints 
on the circumnuclear environment but not uniquely discriminating among 
the competing explanations for the optical-broad-line velocity offsets.

\subsection{Black Hole Mass Constraints}

\begin{figure*}
	\centering
	\includegraphics[width=\textwidth]{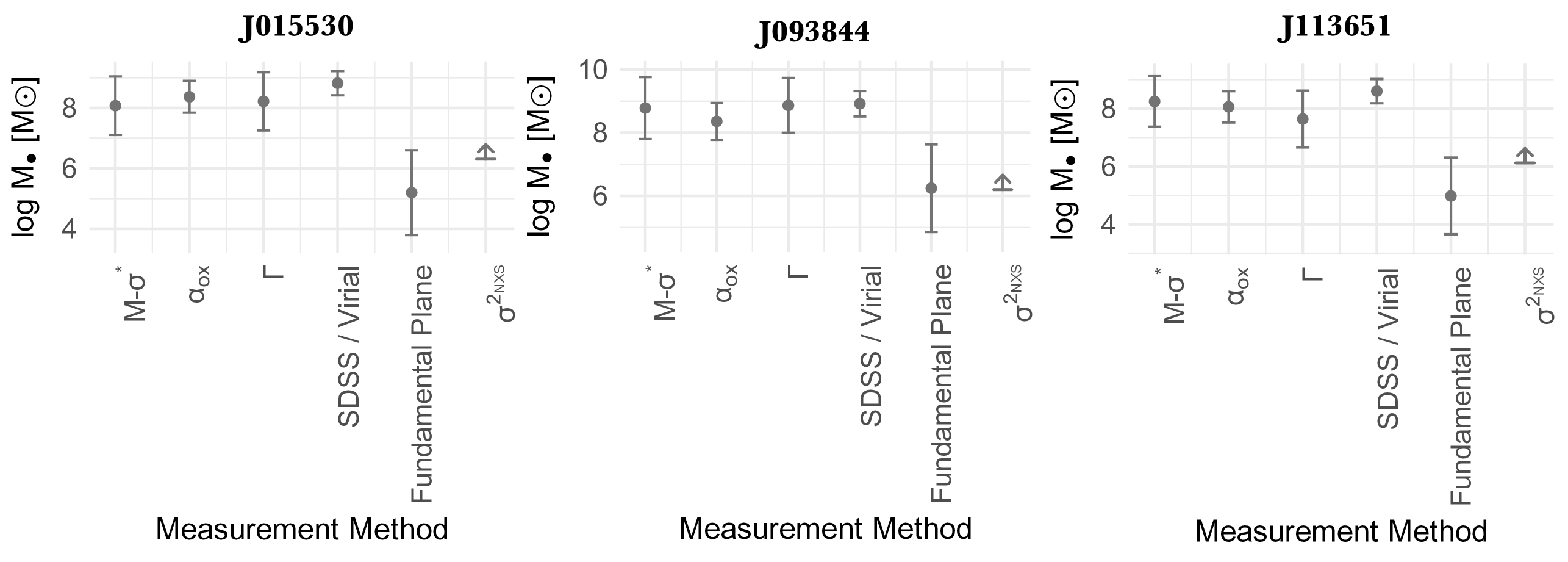}
	\caption{
		Example black hole mass estimates for three quasars in our sample (shown left to right):
		J015530, J093844, and J113651.
		Each panel shows black hole masses inferred using multiple different methods with $1\sigma$ uncertainties.
		The $M_{\bullet}$--$\sigma_{\ast}$ estimate is interpreted as tracing the total black hole mass of the host galaxy, while accretion-based estimators (from $\alpha_{\mathrm{ox}}$ and the X-ray spectral slope $\Gamma$), single-epoch virial masses, and fundamental-plane masses are treated as probing the actively accreting black hole.
		Across these examples, virial masses tend to be systematically higher than accretion-based estimates, while fundamental-plane masses are systematically lower, illustrating the estimator-dependent offsets that motivate the comparison presented in \S\ref{res:mass_comparison}.
	}
	\label{fig:mass_constraints_examples}
\end{figure*}

Our observations allow us to estimate black hole masses through several 
different techniques.  In the binary SMBH interpretation for the 
velocity–offset optical broad lines, accretion-based methods (such as single-epoch 
virial masses, Eddington–ratio estimates, and the black hole fundamental plane) are 
expected to trace the active SMBH responsible for producing the broad emission lines, whereas the 
$M_{\bullet}$–$\sigma_{\ast}$ relation should reflect the combined, total 
black hole mass of both black holes.  In this scenario, the narrow emission-line widths serve as 
proxies for the host-galaxy stellar velocity dispersion, implicitly assuming 
that (i) the narrow-line region is gravitationally dominated by the bulge 
potential and (ii) the classical $M_{\bullet}$–$\sigma_{\ast}$ relation remains 
valid even if the nucleus hosts a close SMBH binary.  

In our sample, the $M_{\bullet}$--$\sigma_{\ast}$ masses are not systematically larger than the accretion-based mass estimates.  
This behavior is illustrated in Figure~\ref{fig:mass_constraints_examples}, which presents example mass estimates for three quasars in our sample: J015530, J093844, and J113651.  
These sources were selected because they exemplify the two dominant systematic trends seen across the full sample: comparatively low black hole masses inferred from the fundamental plane and elevated single-epoch virial mass estimates relative to accretion-based methods.  
The figure is intended as an object-level illustration of the estimator-dependent offsets that motivate the statistical mass comparisons presented in \S\ref{res:mass_comparison}, rather than as a representative summary of the sample. 

Such consistency between $M_{\bullet}$--$\sigma_{\ast}$ and accretion-based masses is expected for single SMBHs and for binary systems with near-equal black hole masses (mass ratios $q \sim 1$).  
In contrast, binary configurations in which the actively accreting black hole is substantially less massive than its companion are expected to produce accretion-based mass estimates that are systematically lower than the $M_{\bullet}$--$\sigma_{\ast}$ values \citep[e.g.,][]{farris+14}.  Our results are less consistent with extreme mass-ratio binaries in which the secondary overwhelmingly dominates the accretion, but remain compatible with more moderate mass ratios or with scenarios in which the $M_{\bullet}$--$\sigma_{\ast}$ relation does not reliably trace the total mass of a close SMBH binary.  The absence of clear evidence for extreme mass-ratio binaries introduces tension with theoretical expectations that large velocity offsets arising from binary orbital motion should preferentially occur in systems with $q \lesssim 10^{-2}$ \citep{kelley+21}. 

As shown in \S\ref{res:mass_comparison}, the single-epoch SDSS virial masses are systematically higher than the accretion-based estimates, a difference that plausibly reflects physical differences in the structure and dynamics of the broad-line region rather than a purely methodological inconsistency.
Single-epoch virial estimators are calibrated for typical, isolated quasars and assume a virialized broad-line region whose size follows the empirical radius--luminosity relation \citep[e.g.,][]{peterson+04,bentz+13,shen+13b}.
In velocity-offset quasars, however, these assumptions may not hold.
Long-term spectroscopic monitoring shows that broad-line profiles in such systems can vary in both shape and centroid on multi-year timescales, indicating that the emitting gas does not always trace a stable, virialized velocity field \citep{runnoe+17}.  If the actively accreting black hole resides in a bound binary, the broad-line region may be truncated or dynamically perturbed by the binary gravitational potential.
In this case, the observed line widths are dominated by gas at smaller radii, while the radius--luminosity relation continues to assign a characteristic BLR size appropriate for an intact, extended BLR.
This mismatch can lead single-epoch virial estimators to overpredict the black hole mass relative to accretion-based methods.

In contrast, the black hole fundamental–plane masses are generally lower than the
other estimates by roughly 1–1.5~dex (see \S\ref{res:mass_comparison}).  While one possibility for systematically lower mass estimates using the black hole fundamental plane is that binary accretion flows may suppress jet production or 
modify the coronal structure giving rise to the X-ray emission, there are a number of more mundane possibilities.

The \citet{gultekin+19} fundamental--plane calibration combines X-ray 
binaries with a heterogeneous AGN sample dominated by low-luminosity 
AGNs and Seyferts, many of which host compact radio cores associated 
with jet activity.  One possibility is that the calibration is anchored by systems in which 
the radio and X-ray emission arise from a coupled jet--accretion flow, 
often interpreted in the context of radiatively inefficient accretion 
states \citep[e.g.,][]{narayan+94,yuan+14}.  In contrast, our quasars are radiatively efficient and predominantly radio 
quiet, with weak compact radio cores and coronal X-ray emission that does 
not follow the jet-linked radio–X-ray coupling that underlies the 
fundamental plane \citep[e.g.,][]{merloni+03,sikora+07}.  This mismatch in accretion state, together with the weaker radio emission 
and the absence of jet-linked X-ray enhancement in radiatively efficient 
systems, can shift quasars below the fundamental plane and bias the 
inferred masses low.  Consistent with this interpretation, the only three radio-loud quasars in our sample (those with $R>10$) are also the only sources for which the fundamental–plane masses either exceed (J095036, J140700) or closely match (J151443) the masses inferred from the other accretion-based estimators, as illustrated in Figure~\ref{fig:mass_comp_panel}.

The radio properties of our quasars also inform the physical 
interpretations for the velocity-offset broad lines.  If the offsets were 
produced by momentum transfer from a relativistic jet, the line-emitting 
regions would likely be viewed at small angles to the jet axis, and 
relativistic beaming would enhance the observed radio luminosities (and thus black hole fundamental plane masses) \citep{blandford+79}.  
The radio weakness of our sample therefore disfavors jet entrainment 
as the primary driver of the broad-line velocity offsets.  A more 
comprehensive analysis of this scenario will be presented in our 
upcoming study of the Very Large Array survey of the full 
\citet{eracleous+12} parent sample (Breiding et al., in prep.).

\subsection{Fe K$\alpha$ Emission Line Diagnostics}
\label{discussion:iron_lines}

\subsubsection{Detection Statistics}
\label{dis:det_stats}

We detect Fe K$\alpha$ emission lines in 6/16 quasars ($\approx$38\%), one of which show evidence of being fully resolved beyond the ACIS instrumental response.  This fraction is substantially below the $\gtrsim$80–90 percent incidence reported for bright, local type 1 AGN observed at high signal to noise \citep[e.g.][]{nandra07,shu+10}, where the narrow Fe K$\alpha$ core is effectively ubiquitous when the data quality is sufficient.  In those high signal-to-noise observations, spectra typically contain tens of thousands of counts, allowing even weak narrow cores with equivalent widths of $\sim40-100$ eV to be routinely detected \citep[e.g.][]{yaqoob+04,shu+10}.

The substantially lower detection fraction in our sample is therefore not surprising given the short exposure times of our observations (see Table~\ref{table:spectral_fitting}).  As demonstrated by the Monte Carlo completeness simulations detailed in Section~\ref{results:ew_completeness}, many of our spectra have insufficient counts to detect narrow core emission at the typical equivalent width levels seen in local AGN, and have even less sensitivity to relativistically broadened Fe K$\alpha$ components.  When propagated across the sample, the completeness simulations predict a number of Fe K$\alpha$ detections fully consistent with the six cases observed.  This indicates that our lower detection rate is driven by the small number of X-ray counts collected rather than by intrinsically weaker or absent iron emission in velocity-offset quasars.

\section{Summary \& Concluding Remarks}
\label{conclusion}

The major goal of this study was to use X-ray diagnostics of AGN accretion to evaluate the competing physical models for quasars that exhibit velocity-offset broad emission lines.  
These models include orbital motion in a gravitationally bound binary SMBH, a recoiling SMBH, bulk motion of the BLR that does not follow the black hole, and other BLR-specific effects.  
To this end, we analyzed new and archival Chandra observations of sixteen quasars with velocity-offset broad optical emission lines drawn from the 88-object parent sample of \citet{eracleous+12}.  
The X-ray properties of the AGN studied here do not uniquely favor a single physical origin for the velocity offsets, but they do provide several key constraints.

Our sample exhibits modestly hard X-ray continua compared to the broader quasar population.  
This behavior may arise from binary-related accretion flows, but it is also consistent with single-SMBH scenarios involving modest neutral absorption, warm absorption, reflection, or intrinsic differences in the X-ray corona.  
Deep follow-up soft-band ($<2$~keV) Chandra grating spectroscopy with the HETGS could determine whether weak absorption contributes to the hard photon indices we observe, while hard-band ($>10$~keV) observations with NuSTAR could assess the role of reflection.  
Binary SMBH accretion flows therefore remain a plausible, but not uniquely favored, explanation for the observed X-ray spectral hardening.

Using several different mass-measurement techniques spanning our X-ray data, VLA radio data, and SDSS optical/UV photometry, we find that the inferred black hole masses are broadly consistent in scale but exhibit clear, method-dependent offsets that carry physical significance.  We used the $M_{\bullet}$--$\sigma_{\ast}$ relation (with narrow-line widths as a proxy for stellar velocity dispersion) to trace the total SMBH mass and tested whether accretion-based mass estimates, which in binary SMBH scenarios may preferentially trace a lower-mass accreting secondary, are systematically lower.  
We did not find a uniform offset of this kind.  Given the large ($\sim$0.5--1 dex) uncertainties affecting all mass estimators, the absence of a clear systematic offset disfavors extreme mass-ratio binaries, while remaining consistent with binaries having more moderate mass ratios.    

In contrast, we find that single-epoch SDSS virial black hole masses are systematically higher than accretion-based mass estimates derived from X-ray diagnostics.  
This offset is unlikely to reflect a purely methodological discrepancy.  Instead, the offset likely reflects physical differences in the structure and dynamics of the BLR in velocity-offset quasars.  
If the actively accreting black hole resides in a bound binary, BLR truncation or dynamical perturbations can violate the assumptions of single-epoch virial estimators and lead to systematically elevated virial masses relative to accretion-based estimates.

We also found systematically lower black hole masses from the fundamental plane of black hole activity compared to the other mass estimators.  
The most parsimonious explanation is that our quasars are predominantly radio-quiet systems with radiatively efficient accretion.  
In this regime, the canonical fundamental-plane relation, which is calibrated primarily on radiatively inefficient, jet-dominated sources, is expected to underpredict black hole masses.  
This result further disfavors relativistic-jet-driven BLR outflows as the dominant explanation for the velocity offsets, since such scenarios would require strong radio emission and relativistic beaming.


\begin{acknowledgements}
Support for this work was provided by the Chandra X-ray Center (CXC) through Cycle 25 Guest Observer Program GO4-25076A (PI: P.  Breiding).  This research has made use of data obtained from the Chandra Data Archive and of software provided by the Chandra X-ray Center in the application packages CIAO and Sherpa.  

We acknowledge use of data from the Sloan Digital Sky Survey.  Funding for the Sloan Digital Sky Survey V has been provided by the Alfred P.  Sloan Foundation, the U.S.  Department of Energy Office of Science, and the Participating Institutions.  SDSS acknowledges support and resources from the Center for High-Performance Computing at the University of Utah.  The SDSS web site is www.sdss.org.

A portion of this work utilized the publicly available data stet VLA/12B-303, collected by the VLA instrument operated by the National Radio Astronomy Observatory (NRAO) (data is available at the following website: http://archive.nrao.edu/archive).  The NRAO is a facility of the National Science Foundation operated under cooperative agreement by Associated Universities, Inc.

\end{acknowledgements}

\bibliography{cxo_binaries_pdb}{}
\bibliographystyle{aasjournal}

\appendix

\section{Scaling Relations}
\label{app:scaling_relations}

In this appendix we list the empirical relations used to estimate Eddington ratios, bolometric luminosities, black hole masses, and host–galaxy velocity dispersions.  For each relation we give the coefficients, their uncertainties, and the corresponding intrinsic scatter we adopt when applying these scaling relations, $\sigma_{\rm int}$.  These relations are applied in Section~\ref{results:mass_estimates}, where all relevant uncertainties are fully propagated.

\subsection{Photon Index and Eddington Ratio}
We estimate the Eddington ratio, $\lambda$, using the correlation with the rest-frame hard X-ray photon index, $\Gamma$,  from \citet{shemmer+08}, adopting their Equation 2.  Their regression for unobscured Type 1 radio-quiet quasars is
\begin{equation}
	\mathrm{log\left(\lambda\right)=\left(0.9\pm0.3\right)\Gamma-\left(2.4\pm0.6\right)}
\end{equation}
with an intrinsic scatter of $\sigma_{\rm int}= 0.35$ in $\log\lambda$.

\subsection{Optical-to-X-ray Color and Eddington Ratio}

As a complementary method for determining the Eddington ratio, $\lambda$, from our X-ray data, we use the correlation between $\alpha_{\rm ox}$ and $\lambda$ from \citet{lusso+10}, where $\alpha_{\rm ox}$ is treated as the independent variable (their Equation 16):
\begin{equation}
	\label{eq:aox_relation}
	\alpha_{\rm ox} = (0.719 \pm 0.127),\log \lambda + (2.124 \pm 0.132).
\end{equation}
We invert this relation to obtain $\lambda$ for each quasar, adopting a scatter of $\sigma_{\rm int}=0.4$ in $\log\lambda$.  Propagating this through the slope in \ref{eq:aox_relation} corresponds to an intrinsic dispersion of $\sigma_{\alpha_{\rm ox}}\approx0.3$, which is consistent with the observed $\alpha_{\rm ox}$ scatter in our sample. 

\subsection{ Bolometric Correction and Eddington Ratio}
To convert hard X-ray luminosity to bolometric luminosity, we adopt the bolometric-correction–Eddington-ratio relation of \citet{lusso+12} which treats the Eddington ratio, $\lambda $, as the independent variable.  This relation is calibrated using 170 Type-1 AGN with black hole masses estimated from broad emission lines, and is given as:
\begin{equation}
	\mathrm{log\left(K_{bol}\right)=\left(0.273\pm0.045\right)log\left(\lambda\right)+\left(1.656\pm0.056\right)},
\end{equation}
with an intrinsic scatter of $\sigma_{\rm int}\!=\!0.27$ in $\log K_{\rm bol}$.  We compute bolometric luminosities using $\mathrm{L_{bol} = K_{bol}L_{2-10~keV}}$, where $\mathrm{L_{2-10~ keV}}$ is determined in the rest frame.

\subsection{Variability and Black Hole Mass}

For each of our quasars with adequately sampled light curves, we constrain black hole mass using the empirical relation between normalized excess variance and black hole mass from \citet{ponti+12}.  We adopt either the 20~ks or the 40~ks calibration from \citet{ponti+12} for the CAIXAvar sample, depending on which is closest to the source’s rest-frame light curve length.  The relation has the form
\begin{equation}
	\label{eq:var_scaling}
	\log \sigma_{\rm NXS}^{2} = \alpha + \beta\,\log M_{\bullet},
\end{equation}.

For the 20~ks calibration, $\alpha=-2.13\pm0.14$, $\beta=-1.24\pm0.12$, and the intrinsic scatter is $\sigma_{\rm int}=0.47$.  For the 40~ks calibration, $\alpha=-2.00\pm0.13$, $\beta=-1.32\pm0.14$, and the intrinsic scatter is $\sigma_{\rm int}=0.49$.  We invert \ref{eq:var_scaling} to obtain 95\% lower limits on the black hole mass based upon our $\sigma^{2}_{NXS}$ upper limits provided in Table~\ref{table:excess_var}.

\subsection{Fundamental Plane of Black Hole Activity}
We also determine black hole masses based on rest-frame 5~GHz radio core and $2-10$~keV X-ray luminosities of our quasars using the \citet{gultekin+19} calibration:
\begin{equation}
	\begin{aligned}
		\label{eq:fun_plane}
		\mathrm{
			\log \left(\frac{M_{\bullet}}{10^{8}M_{\odot}}\right) =  \mu_{0}
			+ \xi_{\mu R}\log\left(\frac{L_{R}}{10^{38}~erg~s^{-1}}\right)} \\ \mathrm{
			+ ~ \xi_{\mu X} \log\left(\frac{L_{X}}{10^{38}~erg~s^{-1}}\right),}
	\end{aligned}
\end{equation}

where $\mu_{0}=0.55 \pm 0.22$, $\xi_{\mu R}=1.09 \pm 0.10$, $\xi_{\mu X}=0.59^{+0.16}_{-0.15}$, and $\sigma_{int}=0.96$.  

\subsection{Stellar Velocity Dispersion and the $M_{\bullet}$--$\sigma_{\ast}$ Relation}

We estimate the host-galaxy stellar velocity dispersion using the width of the narrow [O~III] $\lambda5007$ emission line as a proxy, adopting the empirical calibration of \citet{brotherton+15},
\begin{equation}
	\log \sigma_{\ast} = A + B \,\log {\rm FWHM}({\rm [O\,III]}),
\end{equation}
with $A = -0.12252 \pm 0.0323$, $B = 0.9332 \pm 0.0126$, and an intrinsic scatter of $\sigma_{\rm int}=0.2$ dex. We then compute black hole masses using the \citet{tremaine+02} $M_{\bullet}$--$\sigma_{\ast}$ relation,
\begin{equation}
	\log M_{\bullet} = \alpha
	+ \beta\,
	\log\!\left(\frac{\sigma_{\ast}}{200\,{\rm km\,s^{-1}}}\right),
\end{equation}
with $\alpha = 8.13 \pm 0.06$, $\beta = 4.02 \pm 0.32$, and an adopted intrinsic scatter of $\sigma_{\rm int}=0.3$ dex.

We intentionally adopt the \citet{tremaine+02} calibration as a conservative case. In the high–velocity–dispersion regime relevant for luminous quasars, $M_{\bullet}$--$\sigma_{\ast}$ relations with steeper slopes generally predict larger black hole masses for a given $\sigma_{\ast}$. Using the \citet{tremaine+02} relation therefore yields comparatively lower host-based masses and provides a conservative test of whether accretion-based mass estimates are systematically lower than host-based masses in binary SMBH scenarios.

\end{document}